\documentclass[usenatbib]{mnras}

\usepackage{newtxtext,newtxmath}

\usepackage[T1]{fontenc}

\DeclareRobustCommand{\VAN}[3]{#2}
\let\VANthebibliography\thebibliography
\def\thebibliography{\DeclareRobustCommand{\VAN}[3]{##3}\VANthebibliography}

\DeclareRobustCommand{\DU}[3]{#2}
\let\DUthebibliography\thebibliography
\def\thebibliography{\DeclareRobustCommand{\DU}[3]{##3}\DUthebibliography}

\DeclareRobustCommand{\DE}[3]{#2}
\let\DEthebibliography\thebibliography
\def\thebibliography{\DeclareRobustCommand{\DE}[3]{##3}\DEthebibliography}

\usepackage{graphicx}	
\usepackage{amsmath}	
\usepackage{tabularx}
\usepackage{tablefootnote}

\newcommand{\lya}{\,$\text{Ly}\alpha$\,} 
\newcommand{\kms}{\,$\text{km~s}^{-1}\,$}
\newcommand{\hmpc}{\,$\text{h}^{-1}\text{Mpc}\,$}

\author[S. J. Youles et al.]{
Samantha Youles,$^{1}$\thanks{E-mail: samantha.youles@port.ac.uk}
Julian E.~Bautista,$^{1, 2}$\thanks{E-mail: bautista@cppm.in2p3.fr}
Andreu Font-Ribera,$^{3,4}$
David Bacon,$^{1}$
James Rich,$^{5}$\newauthor
David Brooks,$^{4}$
Tamara M.~Davis,$^{6}$
Kyle Dawson,$^{7}$
Axel de la Macorra,$^{8}$
Govinda Dhungana,$^{9}$
Peter Doel,$^{4}$\newauthor
Kevin Fanning,$^{10}$
Enrique Gaztañaga,$^{11}$
Satya Gontcho A Gontcho,$^{12}$
Alma X.~Gonzalez-Morales,$^{13,14}$\newauthor
Julien Guy,$^{12}$
Klaus Honscheid,$^{15,16}$
Vid Ir\v{s}i\v{c},$^{17,18}$
Robert Kehoe,$^{9}$
David Kirkby,$^{19}$
Theodore Kisner,$^{12}$\newauthor
Martin Landriau,$^{12}$
Laurent Le Guillou,$^{20}$
Michael E.~Levi,$^{12}$
Paul Martini,$^{15,21}$\newauthor
Andrea Muñoz-Gutiérrez,$^{8}$
Nathalie Palanque-Delabrouille,$^{5,12}$
Ignasi P\'erez-R\`afols,$^{3,20}$\newauthor
Claire Poppett,$^{12,22,23}$
César Ram\'irez-P\'erez,$^{3}$
Michael Schubnell,$^{10}$
Gregory Tarl\'{e}$^{10}$
and Michael Walther$^{24,25}$
\\
$^{1}$Institute of Cosmology \& Gravitation, University of Portsmouth, Dennis Sciama Building, Portsmouth, PO1 3FX, UK\\
$^{2}$Aix Marseille Univ, CNRS/IN2P3, CPPM, Marseille, France\\
$^{3}$Institut de F\'{i}sica d’Altes Energies (IFAE), The Barcelona Institute of Science and Technology, Campus UAB, 08193 Bellaterra Barcelona, Spain\\
$^{4}$Department of Physics \& Astronomy, University College London, Gower Street, London, WC1E 6BT, UK\\
$^{5}$IRFU, CEA, Universit\'{e} Paris-Saclay, F-91191 Gif-sur-Yvette, France\\
$^{6}$School of Mathematics and Physics, University of Queensland, 4072, Australia\\
$^{7}$Department of Physics and Astronomy, The University of Utah, 115 South 1400 East, Salt Lake City, UT 84112, USA\\
$^{8}$Instituto de F\'{\i}sica, Universidad Nacional Aut\'{o}noma de M\'{e}xico,  Cd. de M\'{e}xico  C.P. 04510,  M\'{e}xico\\
$^{9}$Department of Physics, Southern Methodist University, 3215 Daniel Avenue, Dallas, TX 75275, USA\\
$^{10}$Department of Physics, University of Michigan, Ann Arbor, MI 48109, USA\\
$^{11}$Institute of Space Sciences, ICE-CSIC, Campus UAB, Carrer de Can Magrans s/n, 08913 Bellaterra, Barcelona, Spain\\
$^{12}$Lawrence Berkeley National Laboratory, 1 Cyclotron Road, Berkeley, CA 94720, USA\\
$^{13}$Consejo Nacional de Ciencia y Tecnolog\'{\i}a, Av. Insurgentes Sur 1582. Colonia Cr\'{e}dito Constructor, Del. Benito Ju\'{a}rez C.P. 03940, M\'{e}xico D.F. M\'{e}xico\\
$^{14}$Departamento de F\'{i}sica, Universidad de Guanajuato - DCI, C.P. 37150, Leon, Guanajuato, M\'{e}xico\\
$^{15}$Center for Cosmology and AstroParticle Physics, The Ohio State University, 191 West Woodruff Avenue, Columbus, OH 43210, USA\\
$^{16}$Department of Physics, The Ohio State University, 191 West Woodruff Avenue, Columbus, OH 43210, USA\\
$^{17}$Kavli Institute for Cosmology, University of Cambridge, Madingley Road, Cambridge CB3 0HA, UK\\
$^{18}$Cavendish Laboratory, University of Cambridge, 19 J. J. Thomson Ave., Cambridge CB3 0HE, UK\\
$^{19}$Department of Physics and Astronomy, University of California, Irvine, 92697, USA\\
$^{20}$Sorbonne Universit\'{e}, CNRS/IN2P3, Laboratoire de Physique Nucl\'{e}aire et de Hautes Energies (LPNHE), FR-75005 Paris, France\\
$^{21}$Department of Astronomy, The Ohio State University, 4055 McPherson Laboratory, 140 W 18th Avenue, Columbus, OH 43210, USA\\
$^{22}$Space Sciences Laboratory, University of California, Berkeley, 7 Gauss Way, Berkeley, CA  94720, USA\\
$^{23}$University of California, Berkeley, 110 Sproul Hall \#5800 Berkeley, CA 94720, USA\\
$^{24}$University Observatory, Faculty of Physics, Ludwig-Maximilians-Universit{\"a}t, Scheinerstr. 1, 81677 M{\"u}nchen, Germany\\
$^{25}$Excellence Cluster ORIGINS, Boltzmannstrasse 2, D-85748 Garching, Germany
}

\title[Quasar redshift errors in Lya forests]{The effect of quasar redshift errors on Lyman-$\alpha$ forest correlation functions}

\date{Accepted XXX. Received YYY; in original form ZZZ}

\pubyear{2022}

\begin{document}
\label{firstpage}
\pagerange{\pageref{firstpage}--\pageref{lastpage}}
\maketitle

\begin{abstract}
Using synthetic Lyman-$\alpha$ forests from the Dark Energy Spectroscopic Instrument (DESI) survey, we present a study of the impact of errors in the estimation of quasar redshift on the Lyman-$\alpha$ correlation functions. Estimates of quasar redshift have large uncertainties of a few hundred \kms due to the broadness of the emission lines and the intrinsic shifts from other emission lines. We inject Gaussian random redshift errors into the mock quasar catalogues, and measure the auto-correlation and the Lyman-$\alpha$-quasar cross-correlation functions. We find a smearing of the BAO feature in the radial direction, but changes in the peak position are negligible. However, we see a significant unphysical correlation for small separations transverse to the line of sight which increases with the amplitude of the redshift errors. We interpret this contamination as a result of the broadening of emission lines in the measured mean continuum, caused by quasar redshift errors, combined with the unrealistically strong clustering of the simulated quasars on small scales.
\end{abstract}




\section{Introduction}
\label{sec:Introduction}


The study of dark energy, as a potential explanation for the accelerated nature of the expansion of the Universe, demands high precision measurements of the expansion rate. These measurements are possible with the use of standard candles or standard rulers, particularly those that are visible out to large distances, 
or equivalently, to early cosmic times. 
\citet{riess_observational_1998} and \citet{ perlmutter_discovery_1998} 
measured the flux and redshift of type-Ia supernovae (SNIa) which are standardizable candles. This enabled the calculation of luminosity distance, $D_{L}$, as a function of redshift, which showed that the expansion of the Universe is accelerating. Since then, both the quantity and quality of recent SNIa data have contributed to the reduction of uncertainties on parameters describing dark energy \citep{scolnic_complete_2018, brout_des_2019}.

The accelerated expansion has been confirmed using a completely 
independent probe: baryon acoustic oscillations (BAO) as a standard 
ruler. These acoustic oscillations in the primordial plasma, prior 
to recombination, left an imprint on the large-scale structure of 
the Universe that corresponds to the size of the sound horizon, 
$r_{d}$, at the drag epoch. This scale manifests as a peak in the
matter-density correlation function at comoving separations 
$\sim$ 147 Mpc, or equivalently, as an oscillatory pattern in the
power spectrum. In the transverse direction, the BAO peak measures
the ratio $D_M(z)/r_d$, where $D_M(z) = (1 + z)D_A(z)$ is the 
comoving angular-diameter distance. In the radial direction, it
determines $D_H(z)/r_d$, where $D_H(z) = c/H(z)$ is the Hubble 
distance. These observables are used to infer the expansion
history of the Universe and derive cosmological parameters of 
the models that describe it.

Since the first BAO measurements \citep{eisenstein_detection_2005, 
cole_2df_2005} several spectroscopic surveys have been built 
with the goal of measuring the BAO scale in the distribution of 
matter. This distribution has been traditionally mapped with galaxies as a tracer. 
As galaxies become relatively faint above redshifts of $z = 1$,
quasars have been used to trace the matter field at those higher redshifts. 
At $z>2$, a new window has been recently opened to observe BAO using 
Lyman-$\alpha$ (\lya) forests, features seen in the spectra 
of quasars, caused by the absorption of light by neutral hydrogen  
\citep{busca_baryon_2013, slosar_measurement_2013, 
kirkby_fitting_2013, delubac_baryon_2015, bautista_measurement_2017,
desainteagatheBaryonAcousticOscillations2019}
and in cross-correlation with quasars \citep{Font-Ribera:2014_xcf, duMasDesBourboux:2017, Blomqvist:2019, du_mas_des_bourboux_completed_2020}.

An accurate estimate of the redshift of quasars is essential for determining the 2-point statistics used for cosmological inference. Errors above several hundred \kms\ on the redshift estimates 
cause the 2-point functions to be smeared in the line-of-sight direction, reducing the precision in BAO measurements (even though this smearing is accounted for in the modelling).
The broad emission line centres in the spectra of quasars are not necessarily good indicators of the host galaxy redshift (also named systemic redshift) which makes it problematic to obtain precise measurements. Due to the complex dynamic of the line-emitting regions in quasars, the broad line centres can be shifted with respect to their expected locations in the quasar rest-frame. 
Using spectra with high signal to noise ratio from 32 epochs in the Sloan Digital Sky Survey Reverberation Mapping project (SDSS-RM), \citet{shen_sloan_2016} calculated the relative velocity shifts between pairs of lines, and between lines and the systemic redshift when available. The systemic redshifts of quasars were obtained when narrow emission lines and/or stellar absorption lines could be observed. Broad high-ionisation lines, such as CIII, are typically blue-shifted by tens or hundreds of \kms\ and may have a strong luminosity dependence.
Conversely the velocity shifts tend to be smaller for low-ionisation 
lines such as MgII, with no luminosity dependence. This study enabled 
the authors to derive  empirical recipes for unbiased estimation of 
redshift with uncertainties based on various lines across a range of 
redshifts.
Template fitting software, such as 
\texttt{redrock}, \footnote{\url{https://github.com/desihub/redrock}} 
uses templates built from a principal component analysis (PCA), 
which cannot fully account for all spectral variations of quasars. 

The eBOSS collaboration \citep{dawson_extended_2016, alam_completed_2021} 
used several different redshift estimators for quasars, 
all included in the official SDSS quasar catalogue \citep{lyke_sloan_2020}
from the Data Release 16 \citep{ahumada16thDataRelease2020}. 
In the clustering analysis of DR16 \lya\ forests,  \citet{du_mas_des_bourboux_completed_2020} performed a detailed study of the impact of different redshift estimators on the BAO parameters, using both real data and mock catalogues. From mocks, they observed that variations in the uncertainty of BAO best-fit parameters of $\sim 0.5\sigma$ with different redshift estimators, was consistent with statistical error. Analogously, photometric redshift uncertainties in galaxy clustering and BAO measurements have been found to reduce the constraining power on the Hubble parameter \citep{Chaves-Montero_2018}.

The Dark Energy Spectroscopic Instrument (DESI, \citealt{desi_collaboration_desi_2016}) has recently begun a five-year programme of observations, in which it will observe three times more $z>2$ quasars than SDSS. They will be used for clustering measurements with the \lya\ forest. DESI is expected to produce roughly 1\% uncertainties in the BAO parameters from these forests.
DESI is a multi-object optical spectrograph that receives light from 5000 optical fibres mounted at the focal plane of the 4-metre class Mayall Telescope, in Arizona, USA. The light of each object is split and dispersed onto three cameras, each one corresponding to blue, red, and infra-red wavelengths. The resolution of spectra is about 2000 -- 3200 in the blue, 3200 -- 4100 in the red and 4100 -- 5000 in the infrared end. For each sky pointing, the exposure time is dynamically tuned to the current observing conditions in order to match the signal to noise ratio obtained for a 1000 second exposure taken in ideal conditions. Spectra are reduced and calibrated with a fully automated spectroscopic pipeline. DESI observes simultaneously different types of targets: emission line galaxies, luminous red galaxies, quasars as tracers and quasars containing \lya\ forests. At the end of its 5-year programme, DESI expects to cover about 14k deg$^2$ of the observable sky. The DESI \lya\ forest survey aims to obtain four observations for more than 800k quasars with redshifts $z>2.1$, corresponding to a density of 60~deg$^{-2}$ (Chaussidon et al. in prep.).
There are three methods of redshift estimation being employed: 
template fitting with \texttt{redrock}
and two machine learning algorithms, quasarNET
\citep{busca_quasarnet:_2018} and SQUEzE 
\citep{perez-rafols_spectroscopic_2020}. Routine visual inspection
is not a feasible option as DESI is expected to observe
$\mathcal{O}(10^{6})$ quasars.

In this work, we quantify the impact of errors in the redshift 
estimates on the clustering of the \lya forest. In particular,
we looked into the impact on BAO parameters derived from the 
\lya auto-correlation and the \lya - quasar 
cross-correlation, using synthetic versions of the completed DESI dataset (five years of observations). 

This paper is organised as follows. 
Section~\ref{sec:data} describes the data sets used in the analysis.  
Section~\ref{sec:methods} describes the methods used in our analysis, 
and the results from the analysis on simulated data are presented in 
Section~\ref{sec:analysis}. 
Section~\ref{sec:model} contains a description of the model for 
the contamination of the correlation functions by redshift errors.
In Section~\ref{sec:discussion}, we discuss the implications of 
our findings and present our conclusions.

In this work, conversion from angular and redshift separations
to physical separation are made using a flat-LCDM model
with $\Omega_{m} = 0.315$.

\section{Synthetic Data Sets}
\label{sec:data}

This work is based solely on results obtained on synthetic sets of \lya\ data 
that mimic properties of the DESI \lya\ survey. In this section we describe
the basic principles behind the production of these synthetic data sets, 
the particularities of DESI data,  some special sets used to test our 
hypotheses, and how we mimic the intrinsic errors in quasar redshift estimates.

\subsection{Procedure for Ly\texorpdfstring{$\boldsymbol{\alpha}$}{alpha} mock creation}
\label{sec:data:ssec:mockcreation}

Synthetic \lya forest datasets are used by spectroscopic surveys such as 
eBOSS or DESI to test the BAO analysis pipeline and evaluate systematic effects. 
They are designed to reproduce the astrophysical and instrumental characteristics 
of the survey data. By having a large number of realisations of synthetic 
surveys, usually a few hundred, we can test systematic effects to high 
precision. 

There are several methods to create mock \lya\ data. N-body hydrodynamical simulations \citep{borde_hydrosim_2014, rossi_hydrosim_2014, walther_nyx_2021, chabanier_AGN_2020} are among the most realistic methods to create forests but are too computationally expensive for producing hundreds of realisations containing volumes as large as the real surveys. Hybrid methods,  \citep[e.g.,][]{peirani_lymas_2014, sorini_sims_2016}, also rely on N-body simulations to calibrate a model used to produce quick lognormal mock catalogues. The method we use in this work to create synthetic data 
is based on correlated Gaussian random fields 
\citep{coles_lognormal_1991, le_goff_simulations_2011, FontRibera_2012_mocks, bautista_mocks_2015}, 
whose correlations follow a given input power spectrum. 
The random field mimics the matter density field in the Universe. 
Quasars are placed via Poisson sampling in regions where the density  
is larger than a given threshold. One dimensional density "skewers" along the
line-of-sight to each quasar are drawn by interpolating the same random field. 
These can then be transformed to a transmitted flux fraction, 
representative of a \lya forest, by adding small-scale fluctuations, 
converting from density to optical depth, 
and adding redshift-space distortions (RSD). 

We used an implementation of the Gaussian method of synthetic \lya forests, named \textsc{LyaCoLoRe} \citep{farr_lyacolore_2020}. \textsc{CoLoRe} mock catalogues use the package CoLoRe\footnote{\url{https://github.com/damonge/CoLoRe}} \citep{Ramirez-Perez:2021} to place quasars in a lognormal density field. Skewers are drawn from each quasar, have small-scale power added to them and these are transformed into transmitted flux fractions using the \textsc{LyaCoLoRe} package. \footnote{\url{https://github.com/igmhub/LyaCoLoRe}} This process produces noiseless transmission fields that have correct clustering properties on scales relevant for BAO studies. \textsc{LyaCoLoRe} stores skewers of metal absorption and a table of high column density systems (HCDs) in its output, which may be optionally added to the \lya\ skewers during subsequent stages of the pipeline. These simulated datasets cover the projected 5-year DESI footprint and contain around one million quasars in the redshift range $z \in [1.8, 3.8]$, of which  700,000 have $z > 2.1$.

\subsection{Simulating DESI spectra of quasars}
\label{sec:data:ssec:lyaproperties}

We generated mock quasar spectra reproducing observational properties of the Dark Energy Spectroscopic Instrument, as described in Section~\ref{sec:Introduction}. All of the mock data sets used in this work have density of forests of 50~deg$^{-2}$ within the full DESI footprint, using the methodology described in \citet{du_mas_des_bourboux_completed_2020}. Prior to Survey Validation, this was the expected density for five years of DESI observations, but this target has since been revised upwards to 60~deg$^{-2}$.

Spectral properties of our mock quasars are simulated with the 
\textsc{desisim} package.\footnote{\url{https://github.com/desihub/desisim}} Continuum templates (for the unabsorbed flux of the quasar with emission lines)\footnote{Sometimes in the literature "continuum" refers to flux without emission lines.} are generated by using functions from the \textsc{simqso} library \citep{2021ascl.soft06008M}, which contains a broad set of tools to generate mock quasar spectra. Templates are composed by a series of broken power-laws with independent Gaussian slope distributions, and a set of emission lines defined by their rest frame wavelength, equivalent width and Gaussian r.m.s. width. The slopes and emission line profile distribution used for the mocks in this work uses a modified version of the  BOSS DR9 model, which includes some emission lines from the composite model of BOSS spectra from Table 4 of \citet{harrisCompositeSpectrumBOSS2016}, and some adjustment of the equivalent widths so that the mean continuum resembles better the one obtained in eBOSS Data Release 14. DESI intend to co-add four observations of \lya\ quasars. We therefore convolve spectra to the instrumental resolution, and add pixel noise corresponding to 4000 second exposures (i.e. four times the nominal exposure time).

Astrophysical contaminants are often included in synthetic realisations of the \lya\ forest.
Metal absorbers were added to mocks for the first time in \citet{bautista_mocks_2015} and were first modelled in the correlation function in \citet{bautista_measurement_2017}, where the change in the BAO peak parameters caused by metal absorption was found to be less than 1\%. 
Similarly, \citet{FontRibera_2012_HCD} simulated the impact of High Column Density absorbers (HCDs), and discussed its impact on the measured correlations.
It is important to model these contaminants in BAO analyses
\citep{du_mas_des_bourboux_completed_2020}, but its impact
is orthogonal to the effect discussed in this paper.
For this reason, we decided not to include them in our mocks.

\newcommand{\sigfog}{\sigma_{v,\rm{FoG}}}
\newcommand{\sigz}{\sigma_{v,z}}

However, we do simulate the impact of non-linear peculiar velocities in the quasar redshifts, a phenomenon known as \textit{Fingers of God} (FoG). Since our mocks are generated from Gaussian fields, we only have access to linear peculiar velocities. Therefore we simulate FoG by applying a random shift to the quasar 
redshifts drawn from a Gaussian distribution with an r.m.s. given by the parameter $\sigfog$. We use by default a value of $\sigfog = 150$ \kms, similar to the 
expected velocity dispersion in halos hosting quasars at $z \sim 2$, but in Section \ref{sec:analysis} we also use mocks with an extreme value of $\sigfog = 500$ \kms.

\subsection{Simulating quasar redshift errors}
\label{sec:data:ssec:zerrors}

Even though we try to capture the diversity of quasar continua in our mocks, fitting algorithms on simulated data often perform better than in real data \citep{farr:2020_quasarnet}.
For this reason, instead of trying to estimate redshifts from our mock spectra we decided to emulate errors in the pipeline redshift estimation.
We add Gaussian random errors to the quasar redshifts in the mock catalogues, using by default an r.m.s. of $\sigz = 500$ \kms.

It is important to highlight a key difference between how we simulate FoG and redshift errors.
Even though they both add random shifts to the quasar redshift, the shift emulating FoG is applied before we generate the quasar continuum.
On the other hand, the shift emulating redshift errors only affects the value of redshift in the quasar catalogue that will be used in the analysis, but it does not change the simulated spectrum. 

\subsection{No-QSO-clustering and no-forest mock data sets}
\label{sec:data:ssec:specialmocks}

As described in Section~\ref{sec:model}, the effect of redshift errors on the \lya\ clustering depends on both the amplitude of the quasar clustering and the smoothing of the mean continuum template. 

In order to test these assumptions, we created two special types of mock data sets in addition to the standard mocks:

\begin{itemize}
    \item \textit{No-QSO-clustering} mock data sets contain quasars randomly distributed in the volume, regardless of the local density, so both the auto-correlation of quasars and the QSO-\lya cross-correlation are zero by construction. The \lya auto-correlation is conserved, since \lya forests are constructed from the correlated underlying density field as in the standard mock sets. 
    
    \item \textit{No-forest} mock data sets were constructed assuming that quasar spectra have no \lya absorption, to allow analysis of the effect without \lya clustering. The quasars are placed at the peaks of the density field as in the standard mock sets, so the QSO auto-correlation is conserved. These data sets are effectively noiseless, having a simulated exposure time of $10^{6}~$ seconds.
\end{itemize}

In Section~\ref{sec:analysis} we show how quasar redshift errors impact the observed clustering of these special mock sets, validating the assumptions of our contamination model.

\section{Methods}
\label{sec:methods}

Current studies of the \lya forest correlation function for the measurement of baryon acoustic oscillations follow a rather simple approach, summarised in this section. Briefly, the steps are:
\begin{itemize}
    \item fit of quasar continua,
    \item estimate of transmission and associated weights,
    \item estimate of correlation functions, 
    \item estimate of distortion and covariance matrices,
    \item fit of the BAO model.
\end{itemize}
Each of these steps can be performed using the publicly available code "Package for Igm Cosmological Correlation Analyses", 
\textsc{picca}.\footnote{\url{https://github.com/igmhub/picca}} \citep{du_mas_des_bourboux_picca_2021} We refer the reader to \citet{du_mas_des_bourboux_completed_2020} 
for the full description of the methodology and its validation. 

\subsection{Continuum fitting}
\label{sec:methods:continuum}

To compute correlations, we use the contrast, $\delta_{q,i}$, of the transmission $F_q(\lambda^{o}_{i})$ at a given observer-frame wavelength $\lambda^{o}_i$ of pixel $i$ and quasar $q$. The transmission contrast is defined as
\begin{equation}
    \delta_{q,i}= \frac{F_q(\lambda^o_i)}{\bar{F}(\lambda^{o}_i)} - 1,
    \label{eq:delta}
\end{equation}
where $\bar{F}$ is the sample's mean transmitted fraction at the absorber redshift, assumed to be only a function of redshift $z$ (or observed wavelength $\lambda^{o}$ if we assume a single transition, such as \lya). 
We can convert between redshift $z$ and observed wavelength $\lambda^{o}$ 
by assuming a given rest-frame wavelength of the absorption. 
In this work, we focus on the \lya\ absorption, for which $\lambda_\alpha = 1216$~\AA.

We use a rest-frame spectral region between the \lya\ and Ly$\beta$ broad emission lines, i.e., $\lambda^{r} \in  [1040,1200]~\text{\AA}$. The observed frame wavelength of the spectra is $\lambda_{\rm obs} \in [3600, 5500] ~\text{\AA}$. 

The transmission $F_{q,i} = F_q(\lambda^o_i)$ is defined as the ratio between the observed flux in a given pixel and quasar, $f_{q,i}$, and the unabsorbed flux level, commonly referred as the {\it continuum}, $C_{q,i}$, such that the contrast can be written as
\begin{equation}
    \delta_{q,i} = \frac{f_{q,i}}{C_{q,i} \bar{F}_i} - 1.
    \label{eq:delta_theory}
\end{equation}
In \citet{du_mas_des_bourboux_completed_2020} and previous studies, the continuum is assumed to be a universal function of wavelength in the rest-frame of the quasar, scaled by a per-quasar linear function of log-wavelength,
\begin{equation}
    \hat{C}_{q,i} = \bar{C}(\lambda^r_i) \left(a_q + b_q \log{\lambda^o_i}\right),
    \label{eq:continuum_model}
\end{equation}
where $\lambda^r_i = \lambda^o_i/(1+z_q)$, $z_q$ is the redshift of the quasar, $a_q$ and $b_q$ are fitted parameters for each quasar, $\bar{C}$ is also referred as the {\it mean continuum}.

In practice, due to noise, spectrograph resolution, and the non-Gaussian nature of the distribution of the transmission, it is hard to break the degeneracy between the mean continuum $\bar{C}$ and the mean transmission $\bar{F}$, and estimate them separately. Therefore, in \citet{du_mas_des_bourboux_completed_2020} they use 
Eq.~\ref{eq:continuum_model} as the model for the product $C_{q,i}\bar{F}_i$ 
in Eq.~\ref{eq:delta_theory}, so we can re-write it as simply:
\begin{equation}
    \hat{\delta}_{q, i} = \frac{f_{q,i}}{\hat{C}_{q, i}} - 1.
    \label{eq:delta_practice}
\end{equation}
Current methods start by assuming a shape for $\bar{C}(\lambda^r)$, 
dividing the observed flux by it, and fitting for $a_q + b_q \log(\lambda_i)$, assuming Gaussian statistics. Once all quasar spectra are fitted, $\bar{C}$ is computed by stacking the ratio $f_i/[a_q+b_q\log(\lambda_i)]$ in the quasar rest-frame. This new mean continuum is then used again to fit for all parameters $a_q$ and $b_q$. 
This process is repeated a few times  until convergence. 
We highlight that for computing $\bar{C}$, an {\it estimate} of the quasar redshift $z_q$ is used. This estimate might be affected by intrinsic biases, scatter or measurement errors.  

If the mean continuum were flat, the redshift smearing would have no effect on the derived template. However, because of broad emission lines in the mean continuum that are further broadened by the redshift errors, the derived template is systematically different from the true mean continuum. As we will see in Section~\ref{sec:model}, this fact combined with quasar clustering modifies the quasar-forest and forest-forest correlations.

\subsection{Correlation functions}
\label{sec:methods:correlations}

The auto-correlation function of \lya\ transmission fluctuations is defined as
\begin{equation}
    \xi^{{\rm Ly}\alpha  \times  {\rm Ly}\alpha}(\vec{r}_A) = \langle \delta(\vec{x}) \delta(\vec{x}+\vec{r}_A) \rangle,
    \label{eq:auto_correlation}
\end{equation}
where $\vec{r}_A$ is the separation vector between two forest 
pixels in the volume, which can be decomposed into separations parallel 
to the line-of-sight, $r_\parallel$, and transverse to the line-of-sight, $r_\perp$.
The cross-correlation between quasars and \lya\ transmission fluctuations 
is defined as
\begin{equation}
    \xi^{{\rm QSO}  \times {\rm Ly}\alpha}(\vec{r}_A) = \langle \delta_Q(\vec{x}) \delta(\vec{x}+\vec{r}_A) \rangle \approx \langle \delta(\vec{x}_Q+\vec{r}_A) \rangle,
    \label{eq:cross_correlation}
\end{equation}
where $\delta_Q(\vec{x})$ is the fluctuation of the number density of quasars.
The approximated formula on the right-hand side is valid under the assumption that quasars are sparse (shot-noise dominated), in which case the cross-correlation is simply the average \lya\ transmission around quasars at positions given by $\vec{x}_Q$ (see Appendix B in \citet{FontRibera_2012_DLA_cross}).
The auto-correlation function between quasars is
$\xi^{{\rm QSO}  \times {\rm QSO}}\left(\vec{r}_A\right) = \langle \delta_Q(\vec{x}) \delta_Q(\vec{x}+\vec{r}_A) \rangle$. The quasar auto-correlation is an important ingredient of our model in Section~\ref{sec:model} and is estimated from the mock catalogues. 

In practice, the ensemble averages in the above definitions of the 
correlation functions are in fact averages over the pairs of objects 
in a given volume. 
Correlation functions are estimated in bins of separation 
$r_\parallel$ and $r_\perp$.
The estimator used for the auto-correlation 
of the \lya\ forest is
\begin{equation}
    \hat{\xi}^{{\rm Ly}\alpha  \times  {\rm Ly}\alpha}_A = 
    \frac{\sum_{(i,j) \in A} w_i w_j \delta_i \delta_j }{\sum_{(i,j) \in A} w_i w_j },
\end{equation}
where $w_i$ is the weight assigned to pixel $i$. The weights used are described in Eq. 4 of \citet{du_mas_des_bourboux_completed_2020}. The sums are over all pairs of pixels for which their separation is within the bounds of bin $A$. 
Analogously, the cross-correlation estimator is
\begin{equation}
    \hat{\xi}^{{\rm QSO}  \times {\rm Ly}\alpha}_A = 
    \frac{\sum_{(q,i) \in A} w_q w_i \delta_i  }{\sum_{(q,i) \in A} w_q w_i  },
\end{equation}
where $q$ indexes quasars. The auto-correlation of quasars is computed with the Landy-Szalay estimator \citep{landy_estimator_1993} defined as
\begin{equation}
    \hat{\xi}_A^{{\rm QSO}  \times {\rm QSO}} = \frac{DD_A - 2DR_A + RR_A}{RR_A},
    \label{eq:landy_szalay}
\end{equation}
where $DD_A = \sum_{(i,j) \in A} w_i w_j / W_{qq}$ is the weighted 
number of quasar pairs in bin $A$, $W_{qq}$ is the weighted total number of available quasar pairs in the volume. A Poisson sample of unclustered points, named randoms, is built following the geometry of the survey. $RR_A$ is the normalised number of random pairs in bin $A$, while $DR_A$ is the number of cross pairs between quasars and randoms. 

The weights $w_i$ assigned to the transmission contrasts $\delta_i$
are the inverse of the total pixel variance, which is assumed to be a 
combination of some intrinsic variance (function of observed wavelength only), and instrumental variance. The intrinsic variance is estimated from the full set of forests. The weights assigned to both forest pixels and quasars also take into account the evolution of their clustering with redshift, so $w_i \propto (1+z_i)^\gamma$. 
The default values used in the standard mocks are the same as the evolution in the eBOSS data, $\gamma = 1.9$ for forest pixels \citep{duMasDesBourboux:2017} and $\gamma = 0.44$ for quasars \citep{du_Mas_des_Bourboux:2019}. Imperfect weighting is not expected to bias the results. For the special mocks, such as no-QSO-clustering or no-forest (see Section~\ref{sec:data:ssec:specialmocks}), the redshift evolution is neglected.

\subsection{Distortion and covariance matrices}
\label{sec:methods:distortion_covariance}

Our continuum fitting procedure uses information from the forest itself, so each $\hat\delta_{q,i}$ is a linear combination of all $\delta_{q,j}$ from the same quasar $q$. When computing the correlation functions, each pair of pixels brings with it contributions from their whole respective forests. This distorts the measured correlation function. 
\citet{bautista_measurement_2017} introduced a method to account for 
this distortion which was also used in subsequent analysis, including
\citet{du_mas_des_bourboux_completed_2020}. 
If the difference between the true $\delta$ and the distorted $\delta$ is a linear function of $\log \lambda$, the distorted correlation is a matrix times the undistorted correlation. This matrix is referred to as the distortion matrix. We compute these distortion matrices for our mock catalogues following the same procedure. 

Real forests contain absorption by elements other than hydrogen, 
such as silicon, nitrogen and iron, which creates spurious correlations. 
None of our mock catalogues contain these elements, as the effect 
is unrelated to the topic treated in this work. 

The covariance matrix of our measurements are estimated by sub-sampling, i.e., 
we divide the footprint into $p$ sub-regions and compute the correlation function 
in each sub-region $\xi_A^p$. The covariance is written as
\begin{equation}
    C_{AB} = \frac{1}{\sum_p W^p_A \sum_p W^p_B} \sum_{p} W^p_A W^p_B \left(\xi^p_A\xi^p_B  - \xi_A \xi_B \right),
    \label{eq:covariance_matrix}
\end{equation}
which assumes that correlations between sub-regions are negligible. 
The number of bins $A$ of our correlation functions is usually larger than 
the number of the sub-regions $p$, so we smooth the covariance matrix by 
assuming that its correlation coefficients are only a function of 
$\Delta r_\parallel = r_\parallel^A - r_\parallel^B$ and 
$\Delta r_\perp = r_\perp^A - r_\perp^B$. We average all correlation 
coefficients that have the same $(\Delta r_\parallel, \Delta r_\perp)$. 

\subsection{Modelling the correlations}
\label{sec:methods:model}

We model the large-scale correlations following the procedure from
\citet{du_mas_des_bourboux_completed_2020}, that we shortly describe here. 

A given correlation function (auto or cross) is defined as a sum of a smooth part
and a BAO peak part:
\begin{equation}
    \xi^{\rm model}_A = \xi^{\rm sm}_A + \xi^{\rm peak}_A(\alpha_\parallel, \alpha_\perp).
    \label{eq:model_xi}
\end{equation}
Only the peak part depends on the BAO dilation parameters $\alpha_\parallel$ and $\alpha_\perp$, defined as
\begin{equation}
    \alpha_{\parallel} = \frac{[D_H(\bar{z})/r_d]}{[D_H(\bar{z})/r_d]_{\rm fid}},~
    \alpha_{\perp} = \frac{[D_M(\bar{z})/r_d]}{[D_M(\bar{z})/r_d]_{\rm fid}},
    \label{eq:alphas}
\end{equation}
where $r_d$ is the comoving size of the sound horizon at drag epoch, 
$D_H(\bar{z}) = c/H(\bar{z})$ is the Hubble distance, 
$D_M(\bar{z}) = (1 + \bar{z})D_{A}(\bar{z})$ is the comoving angular diameter distance assuming a flat universe, and $\bar{z}$ is the mean redshift of the measurement.

The correlation function model between two tracers is the Fourier 
transform of an anisotropic biased power spectrum written as
\begin{equation}
    P^{\rm model}(\vec{k}) = b_i b_j(1+\beta_i \mu^2_k)(1+\beta_j \mu^2_k) P_{\rm QL}(\vec{k}) F_{\rm NL}(\vec{k})G_{\rm bin}(\vec{k}), 
    \label{eq:power_spectrum}
\end{equation}
where $\vec{k}$ is the wavevector, with modulus $k$ and $\mu_k = k_\parallel/k$;
$b_i$ and $\beta_i$ are the linear bias and redshift-space distortions parameters, respectively; $G_{\rm bin}$ accounts for the binning of the correlation function, $F_{\rm NL}$ is a empirical term that accounts for the non-linear effects on small scales, and $P_{\rm QL}$ is the linear matter power spectrum with a empirical anisotropic damping applied to the BAO peak component. The linear matter power spectrum is computed from a Boltzmann solver code, such as \textsc{camb} \citep{lewis_camb_2000}. 

The non-linear term $F_{\rm NL}$ is only included for the cross-correlation in this work since our mock forests are built from Gaussian random fields and do not contain non-linear clustering. 
Given the Gaussian nature of both FoG and redshift errors in the simulated datasets, we use a Gaussian kernel
with width $\sigma_v$ to model both effects in the cross-correlation:
\begin{equation}
    F^{{\rm QSO} \times {\rm Ly}\alpha}_{\rm NL}\left(\vec{k}\right) = 
    \exp{\left[-\frac{\left(k\mu_k \sigma_v\right)^2}{2}\right]} ~.
    \label{eq:non_linear_term_cross}
\end{equation}
This is similar to the Gaussian kernel proposed in
\citet{percival_white_2009}, with a factor of two
difference to take into account that there is only
one quasar field in the cross-correlation.

The final correlation function model accounts for the distortion 
matrix, as discussed in Section~\ref{sec:methods:distortion_covariance}. 
The distorted correlations are written as
\begin{equation}
    \xi^{\rm dist}_A = \sum_{A'} D_{AA'} \xi^{\rm model}_{A'}.
    \label{eq:xi_distorted}
\end{equation}

In this work, we only focus on the effect of redshift errors, so we do not add metals or high-column density systems to the mock catalogues. Therefore, our theoretical model does not have terms that account for these effects. 

\subsection{Fitting the BAO scale}
\label{sec:methods:bao_fits}

BAO fits are made over separations of $r \in [10, 180]$ \hmpc and between directions $\mu \in [0,1]$ and $\mu \in [-1,1]$ for the auto and cross-correlations respectively. The correlation function bin size is 4 \hmpc, corresponding to 1590 separation bins for the auto and 3180 bins for the cross. The joint fit of auto- and cross-correlations uses a total of 4770 measurements.
Four parameters are let free for the fit of the auto-correlations: 
$\alpha_\parallel$, $\alpha_\perp$, $b_{{\rm Ly}\alpha}$ and 
$\beta_{{\rm Ly}\alpha}$. 
For the cross-correlation, an additional parameter $\sigma_v$
(Eq.~\ref{eq:non_linear_term_cross}) 
is also fitted. The same five are also let free for combined fits.
Table~\ref{tab:parameters} summarises the parameters used in this work.

\begin{table}
    \centering
    \caption{Summary of free parameters of the model fitted to the auto, cross and joint correlation functions. The last parameter, $\sigma_v$, is only used for the cross and joint fits.}
    \begin{tabular}{cc}
        Parameter & Description \\
        \hline 
        $\alpha_\parallel$ & Radial BAO dilation parameter (Eq.~\ref{eq:alphas}) \\
        $\alpha_\perp$ & Transverse BAO dilation parameter (Eq.~\ref{eq:alphas}) \\
        $b_{Lya}$ & Linear density bias (Eq.~\ref{eq:power_spectrum}) \\
        $\beta_{Lya}$ & Linear redshift-space distortions parameter (Eq.~\ref{eq:power_spectrum}) \\
        $\sigma_v$ & Lorentzian radial dispersion of velocities (Eq.~\ref{eq:non_linear_term_cross}) \\
    \end{tabular}
    \label{tab:parameters}
\end{table}

During the creation of the \textsc{CoLoRe} mocks, a Gaussian smoothing of 2~\hmpc is applied to the model power spectrum to account for the low resolution of the simulation grid. This adds an extra (isotropic) smoothing to our mocks even before we add FoG or redshift errors. We account for this by fitting Gaussian smoothing terms for the whole model, similar to $G(k)$ in Eq.~\ref{eq:power_spectrum}, while fixing $\sigma_v$ to zero on the fiducial mocks. These values are subsequently used as fixed parameters when fitting mocks that do contain FoG and redshift errors.

\section{Analysis of synthetic data}
\label{sec:analysis}

In this section we present the analysis of the correlations measured from the simulated data described in Section \ref{sec:data}, using the methodology from Section \ref{sec:methods}. 

\begin{figure*}
    \centering
    \includegraphics[width=0.82\textwidth]{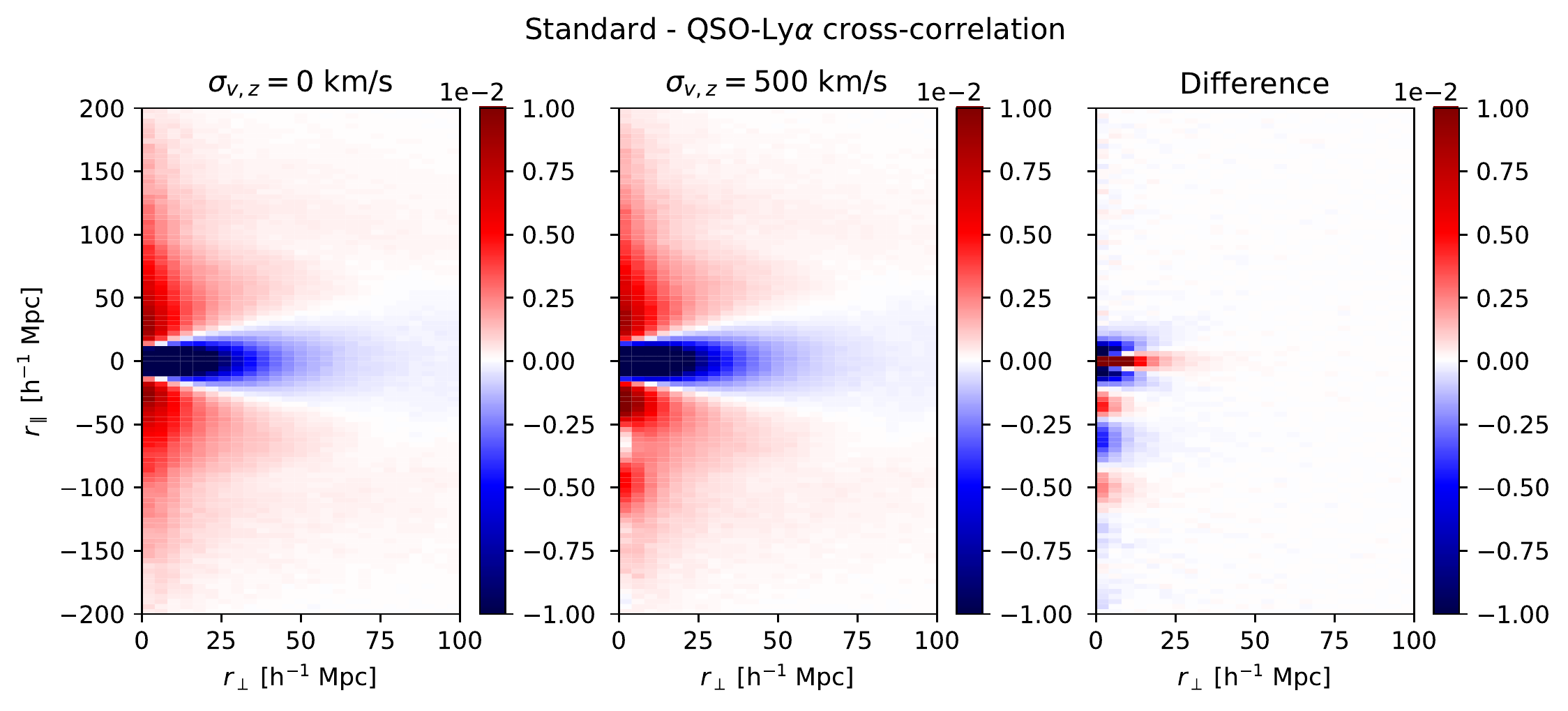}
    \includegraphics[width=0.8\textwidth]{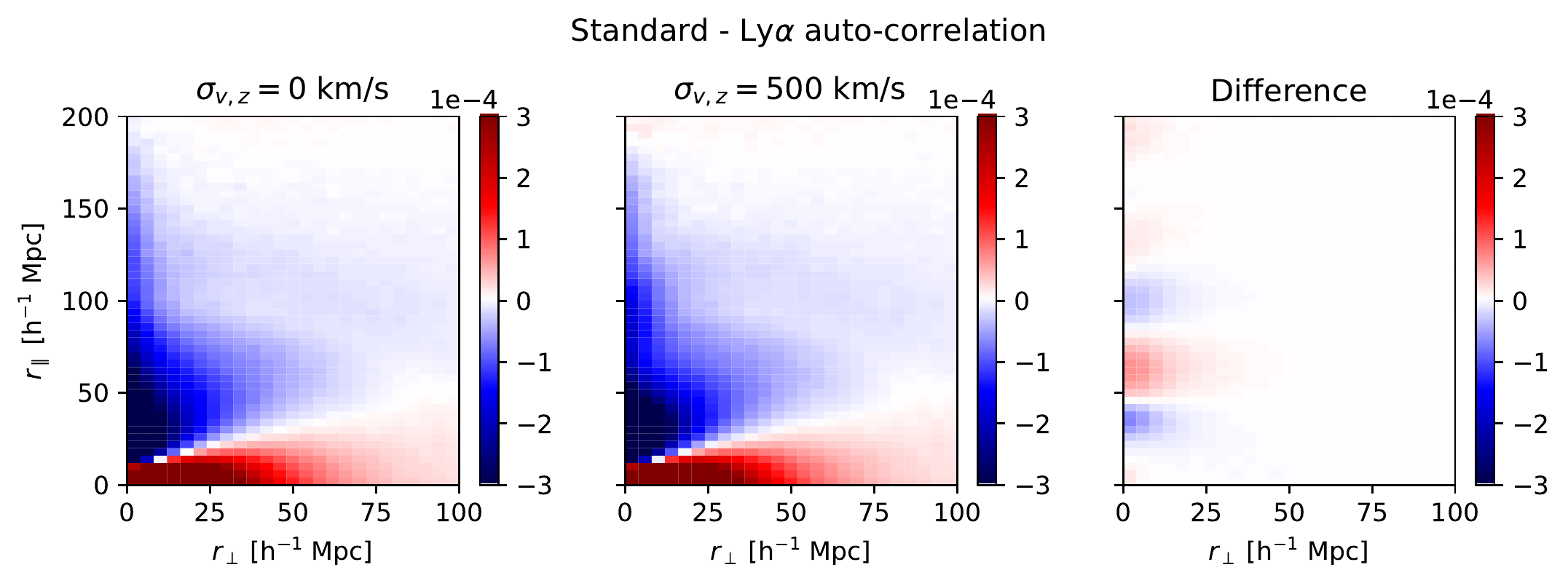}
    \caption{Correlation functions as a function of separations transverse ($r_\perp$) and parallel ($r_\parallel$) to the line of sight in standard mocks. Top panels show the cross-correlation function between \lya\ forests and quasars while bottom panels show the auto-correlation of forests. Left panels show the original correlations while mid panels display correlations with the contamination caused by Gaussian redshift errors of $\sigz = 500$\kms. The right panels show the difference between the left and centre panels, isolating the contamination, which is seen as an oscillating signal at small transverse separations. Note that negative $r_\parallel$ values correspond to \lya\ forest absorption lying between the neighbouring quasar and the observer. This scenario results in a larger auto-correlation between the two quasars than the case where the forest lies behind its neighbouring quasar. }
    \label{fig:xi2d}
\end{figure*}

Figure~\ref{fig:xi2d} presents the estimated correlation functions
versus $(r_\parallel, r_\perp)$ for of the average of ten independent mock realisations. The top panels show the cross-correlation between \lya\ forests and quasars while the bottom ones show the auto-correlation of \lya\ forests. The left panels are mocks without redshift errors ($\sigz =0$\kms) while the central panels contain $\sigz = 500$\kms. The right panels show the difference between central and left panels, isolating the features in the correlation function caused by redshift errors. We can see that, for both the cross- and auto-correlations, these oscillatory features are located at small transverse separations $r_\perp$, decreasing in amplitude as $r_\perp$ increases. Additionally, the cross-correlation signal near zero separations shows the effect of smearing along the line of sight also caused by the redshift errors, and the oscillations occur only for negative $r_\parallel$ values. These values occur when the \lya\ forest pixel is in front of the neighbouring quasar. Where there is a strong cross-correlation, there is also likely to be a strong correlation between the two quasars (as a forest is always in front of its own quasar). For positive $r_\parallel$, where the pixel is behind its neighbouring quasar, its own quasar will be even further behind, so the quasar auto-correlation will be less strong. The asymmetry in the signal, therefore, strongly suggests a dependence on the auto-correlation of quasars.

\begin{figure*}
	\includegraphics[width=0.9\textwidth]{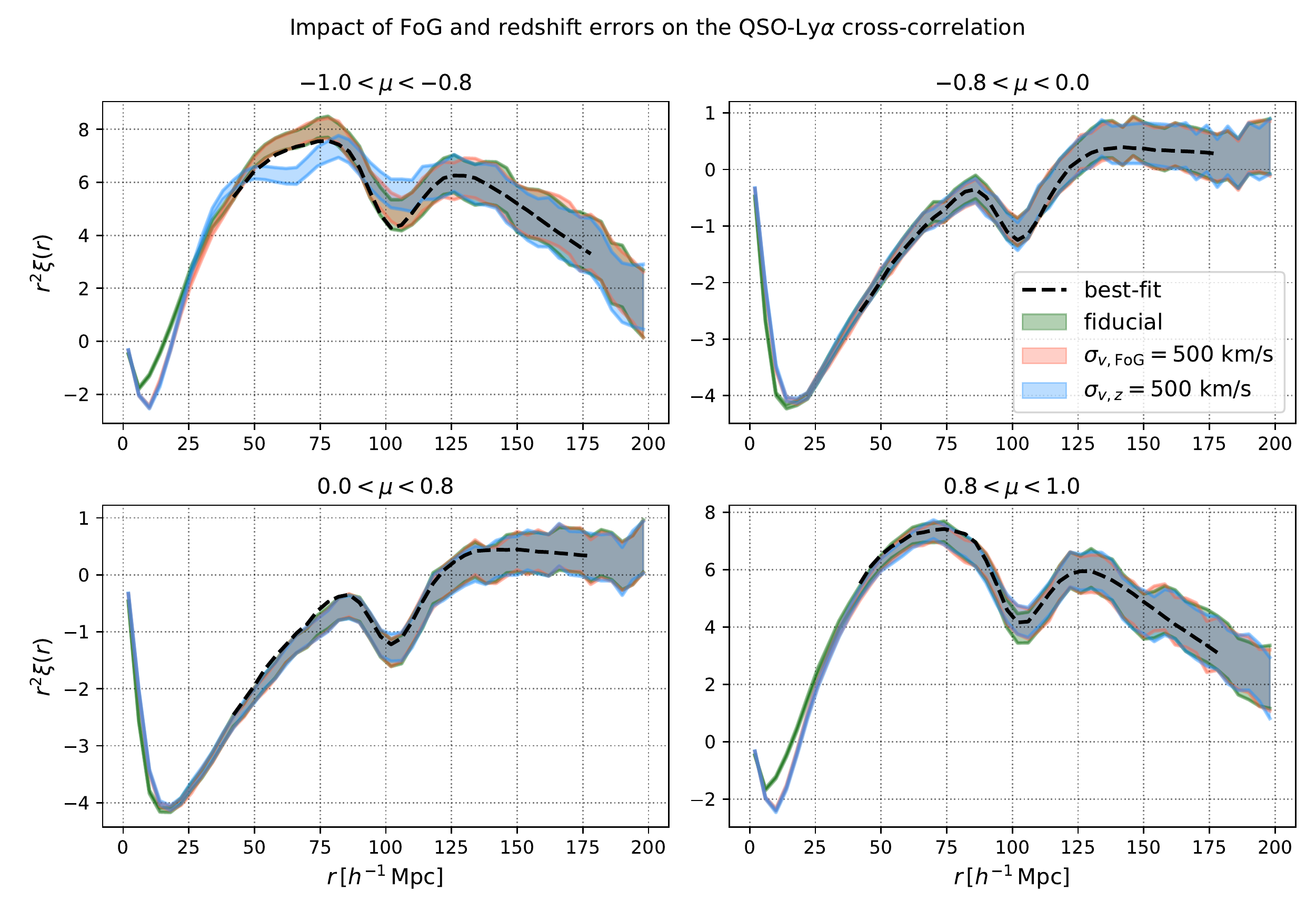}
    \caption {\lya-quasar cross-correlation function from different sets of realisations of the mocks. The green bands show the measurement from the fiducial mocks, with a small value of $\sigfog = 150$ \kms and no redshift errors; the pink bands (indistinguishable from the green in this plot) correspond to the mocks generated with an extreme value of $\sigfog = 500$ \kms; the blue bands shows the results for mocks with large redshift errors of $\sigz =500$ \kms.
    These measurements are computed from the average of 10 realisations of the complete 5-year DESI survey, and the width of the bands corresponds to the scatter between realisations.
    Note that $\mu>0$ ($\mu<0$) corresponds to configurations where the \lya\ pixels are behind (in front of) the neighbouring quasar.}
    \label{fig:wedges_cross}
\end{figure*}

\begin{figure*}
	\includegraphics[width=0.9\textwidth]{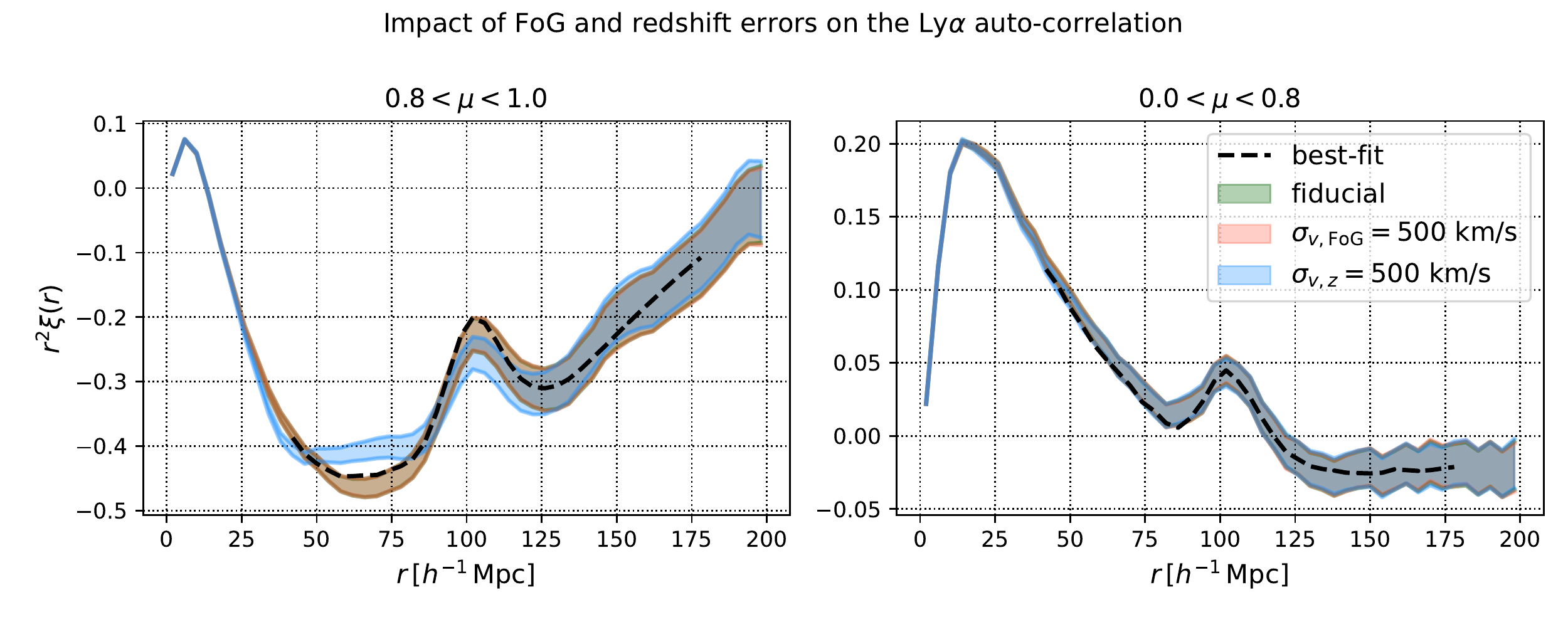}
    \caption{Same as Fig.~\ref{fig:wedges_cross} but for the \lya auto-correlation. As expected, Fingers of God do not impact the auto-correlation. However, 
    the impact of redshift errors $\sigz$ is clearly visible. }
    \label{fig:wedges_auto}
\end{figure*}

Figures~\ref{fig:wedges_cross} and \ref{fig:wedges_auto} show respectively the cross-correlation of the \lya forest with quasars and the auto-correlation of \lya forests, measured from different sets of mocks. Different panels in these figures show different \emph{wedges} of the correlations, i.e., averages over ranges of 
$\mu = r_\parallel/r$. The green bands correspond to the average of ten fiducial 
realisations (with different cosmic variance and instrumental noise), using the standard value of $\sigfog = 150$~\kms and ignoring redshift errors ($\sigz = 0$);
the pink bands are from ten mocks with a larger value for $\sigfog = 500$~\kms; the blue bands show measurements on the fiducial mocks, after adding redshift errors to the quasar catalogues ($\sigz = 500$~\kms). The width of the bands correspond to the standard deviation between the ten realisations of each dataset, i.e., an estimate of the errors for one realisation. The dashed black lines show the best-fit model obtained when analysing the fiducial set of mocks.

\subsection{Impact of non-linear peculiar velocities (FoG)}

Adding FoG to the quasar spectra has a negligible effect in the \lya auto-correlation, making it impossible to distinguish the green and pink bands in Fig.~\ref{fig:wedges_auto}. This is expected since the redshifts of absorption lines do not depend on the quasar redshift. 
On the other hand, the cross-correlation with quasars is smoothed out by the random shifts of the quasar position along the line of sight.
This causes small differences on BAO scales, but the impact is 
clearly seen on scales below 25\hmpc in Figure~\ref{fig:wedges_cross}. 

The impact of FoG on the best-fit parameters from the BAO fits can be seen in Table \ref{tab:bao_fog_sigz}, where the results from the average of ten realisations are compared. Here again FoG have a negligible impact on the \lya auto-correlation, but the uncertainties on $\alpha_\parallel$ from the cross-correlation are 15\% larger (from $0.42$ to $0.48\%$).

\begin{table*}
	\centering
	\caption{Best-fit parameters from ten combined realisations of standard mocks, with simulated Fingers of God velocities of $\sigfog = 150$ or $500$ \kms, and simulated redshift errors of $\sigma_{v,z} = 0$ or $500$ \kms.
	The fourth and fifth columns show the BAO parameters. 
	The last column $\sigma_{v}$ shows the best-fit value of the parameter describing the line-of-sight (Gaussian) smoothing affecting the cross-correlation, which includes FoG and other sources of error on the redshift measurement. The model used to fit the correlations does not
	attempt to account for the new effect discussed in this work. 
	The value of $\sigma_v$ can be
	expressed in units of \hmpc by simply dividing by 
	$H_0 = 100 h$ km s$^{-1}$ Mpc$^{-1}$.
    }
	\label{tab:bao_fog_sigz}
	\begin{tabular}{lcc|cccc}
	\hline
	\hline
		Data Set &  $\sigma_{v,\rm{FoG}}~[{\rm km/s}]$  &  $\sigma_{v,z}~[{\rm km/s}] $ & $(\alpha_{\parallel}-1) \times 10^{3}$ & $(\alpha_{\perp}-1) \times 10^{3}$ & $\sigma_{v}~[{\rm km/s}] $ \\
		\hline
 
Ly$\alpha$ x Ly$\alpha$ & 150 & 0 & 1.9 $\pm$ 4.6 &4.6 $\pm$ 5.9 & - &  \\
Ly$\alpha$ x Ly$\alpha$ & 500 & 0 & 1.5 $\pm$ 4.6 &4.8 $\pm$ 5.9 & - &  \\
Ly$\alpha$ x Ly$\alpha$ & 150 & 500 & -3.8 $\pm$ 5.1 &7.7 $\pm$ 6.4 & - &  \\
 \hline
Ly$\alpha$ x QSO & 150 & 0 & -3.2 $\pm$ 4.2 &-0.1 $\pm$ 4.5 &0 $\pm$ 6 & \\
Ly$\alpha$ x QSO & 500 & 0 & -3.5 $\pm$ 4.8 &-1.0 $\pm$ 4.6 &575 $\pm$ 3 & \\
Ly$\alpha$ x QSO & 150 & 500 & -5.0 $\pm$ 5.2 &1.8 $\pm$ 4.8 &612 $\pm$ 3 & \\
 \hline
Combined & 150 & 0 & -1.2 $\pm$ 3.2 &2.1 $\pm$ 3.6 &0 $\pm$ 4 & \\
Combined & 500 & 0 & -0.7 $\pm$ 3.3 &1.1 $\pm$ 3.6 &561 $\pm$ 2 & \\
Combined & 150 & 500 & -4.4 $\pm$ 3.6 &4.1 $\pm$ 3.9 &588 $\pm$ 2 & \\
 \hline
	\end{tabular}
\end{table*}

\subsection{Impact of quasar redshift errors}

The blue bands in Figures~\ref{fig:wedges_cross} and \ref{fig:wedges_auto}
show the impact of redshift errors.
As discussed in Section \ref{sec:data:ssec:zerrors}, we add these 
redshift errors to the catalogues after the quasar spectra have been
simulated, i.e., the redshifts listed are different than the redshifts 
that were used to generate the quasar continua.

\begin{figure*}
	\includegraphics[width=\textwidth]{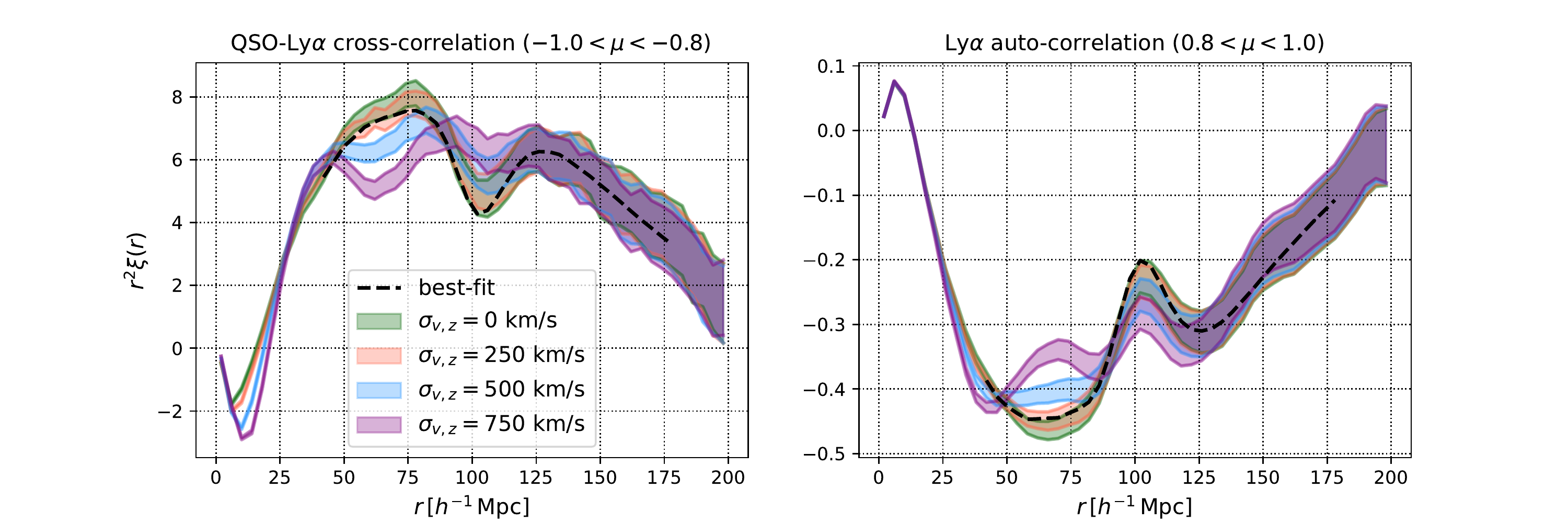}
    \caption{QSO-\lya cross-correlation (left)  and \lya auto-correlation (right)  
    measured from stacks of 10 mock realisations, each stack with different 
    values of Gaussian redshift 
    errors: $\sigz=0$, $250$, $500$, and $750$ \kms.
    We only show the line-of-sight wedges that show spurious correlations 
    caused by redshift errors. These significantly alter the broad-band shape, which will affect the fitting of the correlation functions.} 
    \label{fig:wedges_sigz}
\end{figure*}

For correlations that are not along the line of sight ($| \mu | < 0.8$), the impact of FoG and redshift errors are indistinguishable: the cross-correlation is smoothed on small scales, and the auto-correlation is not affected.
However, clear differences appear in the $-1 < \mu < -0.8$ cross-correlation wedge (top left panel of Fig.~\ref{fig:wedges_cross}) and more surprisingly in the $0.8 < \mu < 1$ auto-correlation wedge (left panel of Fig.~\ref{fig:wedges_auto}).
As shown in Table \ref{tab:bao_fog_sigz}, redshift errors seem to degrade the BAO performance not only in the cross-correlations, but also in the \lya auto-correlation. 

We believe that this is the first time that these features are detected and discussed.

In Fig.~\ref{fig:wedges_sigz} we show the contaminated wedges for realisations with different redshift errors $\sigma_{v,z} = 0$, $250$, 
$500$ and 750~\kms. 
It is clear that the amplitude of the spurious correlations grows monotonically with the amplitude of the redshift errors, and importantly, changes the broad-band shape of the correlation functions which has an impact on the fitting procedures.

Table~\ref{tab:bao_vs_sigz} shows results of BAO fits to mock realisations with increasing values of $\sigz$. We report the average best-fit parameters of ten 
independent realisations for each case. The reported errors are therefore $\sqrt{10}\sim 3.16$ smaller than the expected errors of the full 5-year DESI survey. The best-fit dilation parameters from the auto-correlation function do not present significant changes when increasing $\sigz$, while uncertainties do increase slightly for larger redshift errors. 
The best-fit $\sigma_v$ values correlate well with the input $\sigz$, but are not in agreement, likely due to the fact that $\sigma_v$ also accounts for FoG, or that our model (Eq.~\ref{eq:non_linear_term_cross}) is not necessarily a good match to the Gaussian errors added to mocks. 

These results show that BAO measurements are not biased, even when considering large values for quasar redshift errors $\sigz$. Only the estimated errors on $\alpha_{\parallel}$ and $\alpha_{\perp}$ are increased for larger values of $\sigz$, likely due to the lack of modelling of the effect caused by redshift errors. Properly accounting for these features in the model could potentially help to reduce these errors. In the next section we discuss a potential way to achieve this goal.

\begin{table*}
	\centering
	\caption{Average best-fit parameters of 10 independent realisations of standard mocks, with simulated redshift errors of $\sigma_{v,z} = 0$, 250, 500 and 750 \kms. All these mocks contain simulated FoG with $\sigfog = 150$~\kms.
	The reported errors are the estimated errors of the mean of 10 realisations so they are $\sqrt{10}$ smaller than the expected errors from the full 5-year
	DESI survey. The model used to fit the correlations does not attempt to account for the new effect discussed in this work. The value of $\sigma_v$ output by the fitter originally in units of \hmpc and was converted to \kms by multiplying it by $H(z)/(1+z) = 103.9 h$~km~s$^{-1}$~Mpc$^{-1}$ for $z=2.3$.}
	\label{tab:bao_vs_sigz}
	\begin{tabular}{lc|ccc}
	\hline
		Data Set & $\sigma_{v,z}~[{\rm km/s}] $ & $(\alpha_{\parallel}-1) \times 10^{3}$ & $(\alpha_{\perp}-1) \times 10^{3}$ & $\sigma_{v}~[{\rm km/s}] $ 
		\\

\hline
Ly$\alpha$ x Ly$\alpha$ & 0 & $1.8 \pm 3.6 $ & $8.0 \pm 5.3 $ & -  \\
Ly$\alpha$ x Ly$\alpha$ & 250 & $1.0 \pm 3.7 $ & $9.5 \pm 5.3 $ & -  \\
Ly$\alpha$ x Ly$\alpha$ & 500 & $1.6 \pm 4.1 $ & $9.4 \pm 5.6 $ & -  \\
Ly$\alpha$ x Ly$\alpha$ & 750 & $5.4 \pm 4.8 $ & $5.7 \pm 6.0 $ & -  \\
\hline
Ly$\alpha$ x QSO & 0 & $-4.3 \pm 3.6 $ & $1.0 \pm 4.2 $ & $0 \pm 5 $ \\
Ly$\alpha$ x QSO & 250 & $-6.0 \pm 3.8 $ & $2.7 \pm 4.1 $ & $188 \pm 6 $ \\
Ly$\alpha$ x QSO & 500 & $-8.0 \pm 4.5 $ & $1.3 \pm 4.4 $ & $646 \pm 3 $ \\
Ly$\alpha$ x QSO & 750 & $0.7 \pm 6.2 $ & $-3.3 \pm 5.0 $ & $1012 \pm 3 $ \\
\hline
Combined & 0 & $-0.9 \pm 2.6 $ & $3.7 \pm 3.3 $ & $0 \pm 4 $ \\
Combined & 250 & $-1.9 \pm 2.7 $ & $5.1 \pm 3.3 $ & $150 \pm 6 $ \\
Combined & 500 & $-2.0 \pm 3.1 $ & $4.2 \pm 3.5 $ & $624 \pm 2 $ \\
Combined & 750 & $4.3 \pm 3.8 $ & $0.4 \pm 3.8 $ & $987 \pm 2 $ \\
\hline
	\end{tabular}
\end{table*}

\section{Model for the contamination}
\label{sec:model}

In this section we describe theoretically the expected impact of redshift 
errors in the correlation functions and we will show that the two main 
elements that are responsible for the contamination are:
\begin{enumerate}
    \item a systematic error in the mean continuum estimate, $C_{q,i}$ (Eq.~\ref{eq:continuum_model}), and 
    \item a non-zero quasar-quasar correlation function.
\end{enumerate}

\begin{figure}
	\includegraphics[width=\columnwidth]{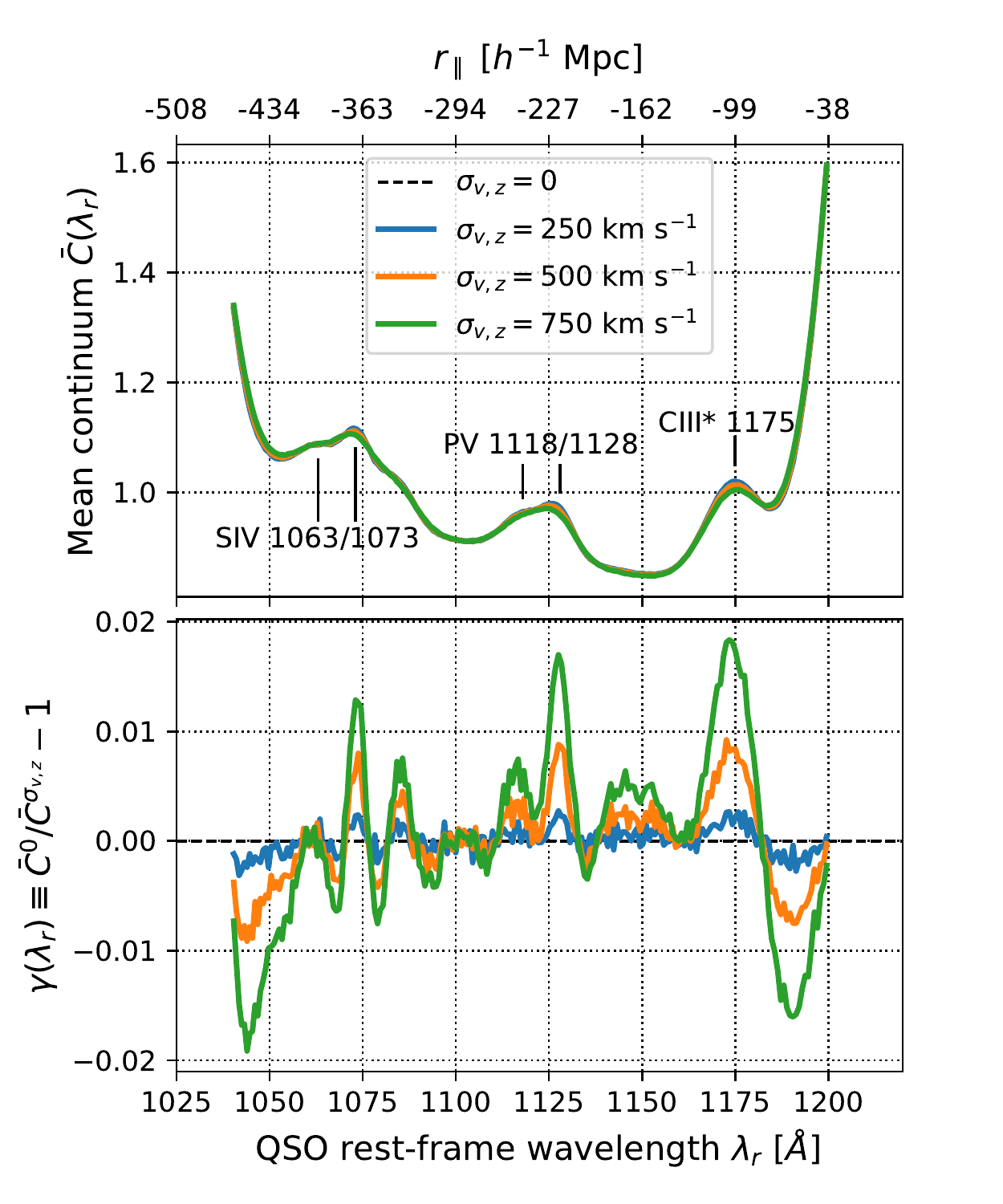}
    \caption{Effect of redshift errors on the mean continuum. The top panel shows the mean continua for no-forest mocks without errors, and with increasingly large redshift errors: $\sigz = 0, 250, 500$ and $750~ \text{km s}^{-1}$. The positions of some of the strongest emission lines are annotated. The lower panel shows the $\gamma(\lambda_{r})$ function defined in Eq.~\ref{eq:mean_continuum_biased}. The top axis on the top panel shows the radial separations between pixels and the quasar, assuming
    $z_q = 2.3$.
    }
    \label{fig:mean_cont}
\end{figure}

The first effect of redshift errors, if assumed to be reasonably distributed randomly around some central value (which is not necessarily the true value), is to smooth the estimate of the mean continuum $\bar{C}(\lambda^r)$. 
Emission lines present in the mean continuum are smeared. 
Figure~\ref{fig:mean_cont} shows how the estimated mean continuum from 
mock catalogues is modified when increasing redshift errors. Known emission lines in the forest region include SIV 1063, 1073, FeII 1082, OIII 1084, PV 1118, 1128, and CIII* 1175. The smoothing of these lines in the continuum is clearly visible in the bottom panel, which shows the relative difference between mean continua with and without errors. Redshift errors of this magnitude introduce biases in the mean continuum of the order of half a percent. These biases have the typical shape of the subtraction of two line profiles with different widths. 

\begin{figure}
	\centering
	\includegraphics[width=\columnwidth]{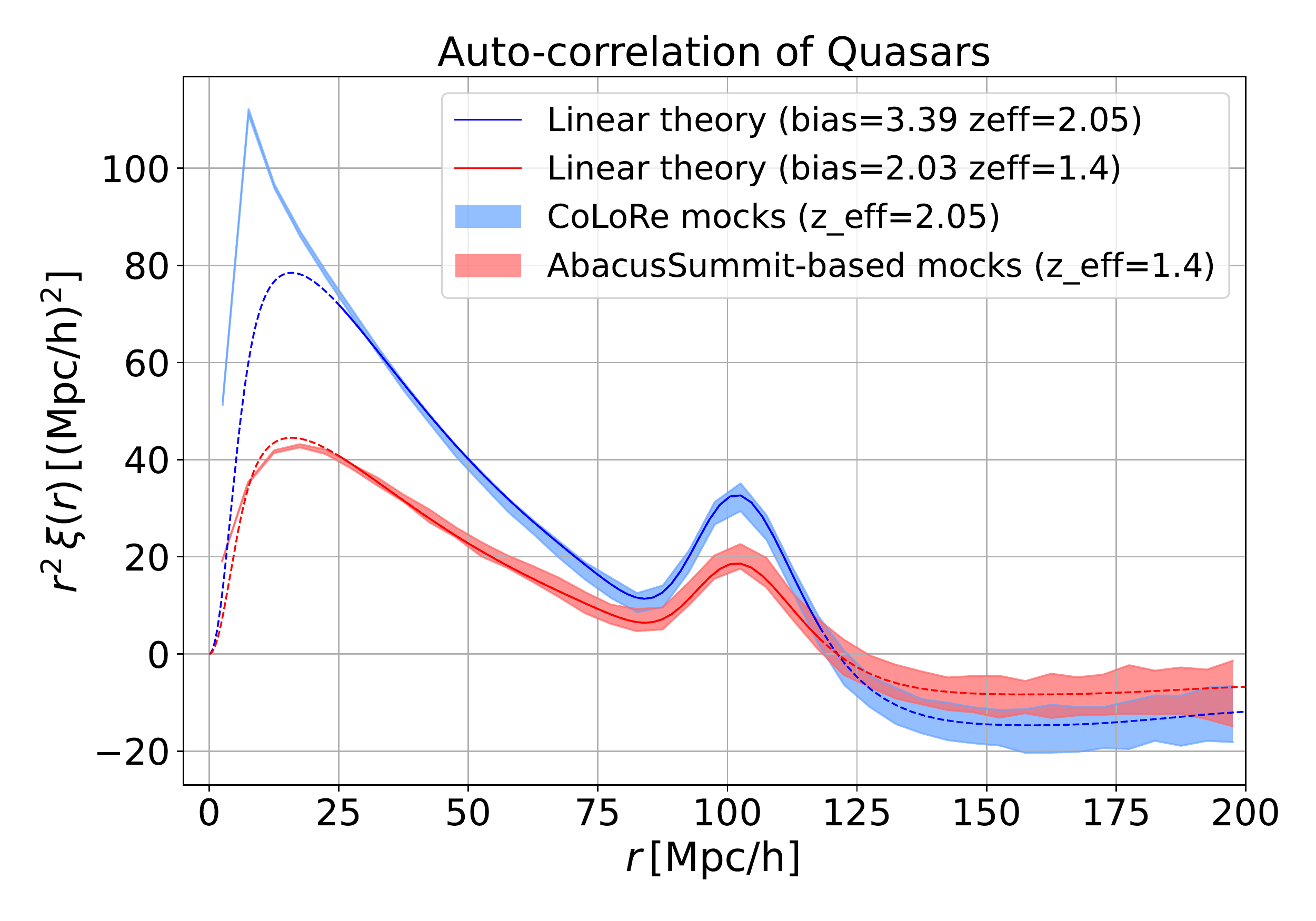}
    \caption{Correlation function of quasars in real space (no RSDs) in CoLoRe boxes (at z=2.05), compared with AbacusSummit-based mocks (Alam et al., in prep.) at z=1.4, with the best-fit linear model, fitted on scales larger than 25 $\text{h}^{-1}\text{Mpc}$. The deviations from linear theory in the CoLoRe mocks are significantly stronger than in AbacusSummit mocks. The high clustering at small scales makes these mocks particularly sensitive to the contamination from redshift errors.}
    \label{fig:auto_qq}
\end{figure}

If we assume a systematic error $\gamma(\lambda^r)$ in the estimate of the mean continuum relative to the true mean continuum $\bar{C}_q(\lambda^r)$, 
as shown in the bottom panel of Figure~\ref{fig:mean_cont}, such that
\begin{equation}
    \bar{C}^*_q(\lambda^r_i) = \bar{C}_q(\lambda^r_i) \left[1+\gamma(\lambda^r_i)\right],
    \label{eq:mean_continuum_biased}
\end{equation}
then the biased transmission contrast can be written from 
Eq.~\ref{eq:delta_practice} as
\begin{equation}
    \hat{\delta}^*_i = \frac{(1+\hat{\delta}_i)}{(1+\gamma_i)} - 1 
            \approx \hat{\delta}_i - \gamma_i, 
    \label{eq:delta_biased}
\end{equation}
where second-order terms have been neglected in the approximation. 

When computing correlations (auto- or cross- with quasars) with this biased contrast field, we perform averages over the distribution of quasars and observed/rest-frame wavelengths. Given that quasars are not uniformly distributed over the volume, i.e., they are clustered, the average of $\gamma$ will not average to zero. This gives the intuition for the second element in our model: the clustering of the quasars. 

\subsection{Impact on quasar-Ly\texorpdfstring{$\boldsymbol{\alpha}$}{alpha} cross-correlation}
\label{sec:model:ssec:qsolya}

We consider two quasars placed at redshifts $z_{q1}$ and $z_{q 2}$. Their separation vector is $\vec{r}_Q$. We will study the cross-correlation between quasar and the forest with a given separation vector $\vec{r}_A$ (corresponding to a given separation bin $A$ in a binned estimation of the correlation function), that can be decomposed into parallel separation $r^A_\parallel$ and transverse separation $r^A_\perp$. Note that given the small angles involved, we can 
assume that the transverse component of $\vec{r}_Q$ is $r^Q_\perp = r^A_\perp$.
For a given $r^A_\parallel$ and $z_{q1}$, the forest pixel will be at the observed wavelength $\lambda^{o}_{A}$ given by $\lambda^o_A=\lambda_\alpha(1+z_A)$, where 
$z_{A} = z_{q1} +  r^A_\parallel /D_H(z_{q1})$. 
The corresponding quasar rest-frame wavelength of the same pixel 
is given by $\lambda^r_A = \lambda^o_A/(1+z_{q2})$. 

The cross-correlation $\hat{\xi}_A$ is the stack of all $\hat{\delta}^*_i$ 
(Eq.~\ref{eq:delta_biased}) lying at separation $\vec{r}_A$ from the quasars. 
\begin{equation}
    \hat{\xi}^*_{A} = \langle \hat{\delta}^* \rangle_{A} 
        = \langle \hat{\delta} \rangle_{A} - \langle \gamma \rangle_{A} 
        = \hat{\xi}_{A} - \langle \gamma \rangle_{A}
	\label{eq:mean_xi_A}
\end{equation}
The average $\langle \gamma \rangle_A$ can be written as a double integral over $z_{q1}$ and $z_{q2}$. The first integral is weighted by the distribution of quasars versus redshift $P(z_{q1})$, while the second is weighted by the distribution of quasars around $q1$ at a given transverse separation $r_\perp^Q = r_\perp^A$, which we denote $P(z_{q2} \mid z_{q1}, r^Q_\perp)$. The excess probability of finding a quasar $q_2$ at a distance $\vec{r}_Q$ from quasar $q_1$ is given by the quasar-quasar correlation function $\xi^{qq}(\vec{r}_Q)$, so $P(z_{q2} \mid z_{q1}, r_\perp)=  P(z_{q2}) \left[1+\xi^{qq}(\vec{r}_Q)\right]$.

\begin{equation}
\begin{aligned}[t]
    \langle \gamma \rangle_{A} &= \int {\rm d} z_{q1} P(z_{q1}) \int {\rm d} z_{q2}  
        P(z_{q2} \mid z_{q1}, r_\perp) ~ \gamma(\lambda^r_A) \\
        &=  \int {\rm d} z_{q1} P^2(z_{q1}) \int {\rm d} z_{q2}  
        \left[1+\xi^{qq}(\vec{r}_Q)\right] ~ \gamma(\lambda^r_A).
	\label{eq:gamma_A}
\end{aligned}
\end{equation}
Note that for a given set of $z_{q1}$, $z_{q2}$ and $\vec{r}_A$, there is only a single possible value for $\lambda^r_A$. Also, $\gamma(\lambda^r)$ is only defined in the forest region, usually between 1040 and 1200\AA.

Given that $P(z_q)$ is a slow-varying function of $z_q$ compared to
$\gamma(\lambda_A^r)$, the first term in the square brackets is washed out by the integrals over $z_{q}$, leaving
\begin{equation}
    \langle \gamma\rangle_A = \int {\rm d} z_{q1} P^2(z_{q1}) \int {\rm d} z_{q2} ~ \xi^{qq}(\vec{r}_Q) ~ \gamma(\lambda^r_A).
	\label{eq:contamination_cross}
\end{equation}

The level of contamination therefore depends on the clustering of quasars $\xi^{qq}$ and on the amplitude of $\gamma$, which is determined by the amplitude of redshift errors. In mock catalogues, we can estimate $\xi^{qq}$ to use in the calculation of the contamination.

Figure~\ref{fig:auto_qq} shows the auto-correlation function of quasars from CoLoRe and AbacusSummit simulations \citep{Garrison_AbacusSummit:2021, Maksimova_AbacusSummit:2021, Hadzhiyska_AbacusSummit:2021, Bose_AbacusSummit:2021}, without redshift space distortions. The AbacusSummit mocks are tailored for quasars from DESI SV (Survey Validation) (Alam et al., in prep). It can be seen that the small-scale clustering in the CoLoRe mocks is significantly stronger than in AbacusSummit. This makes the CoLoRe mocks particularly suitable for testing this model.

Eq.~\ref{eq:contamination_cross} was used to calculate a model for the contamination in the cross-correlation function introduced by redshift errors. 
This equation requires two inputs, first, the systematic bias in the mean continuum estimate and, second, the quasar-quasar correlation function. We compare our model with measurements performed on mock catalogues in the following section.

\begin{figure*}
    \centering
    \includegraphics[width=0.45\textwidth]{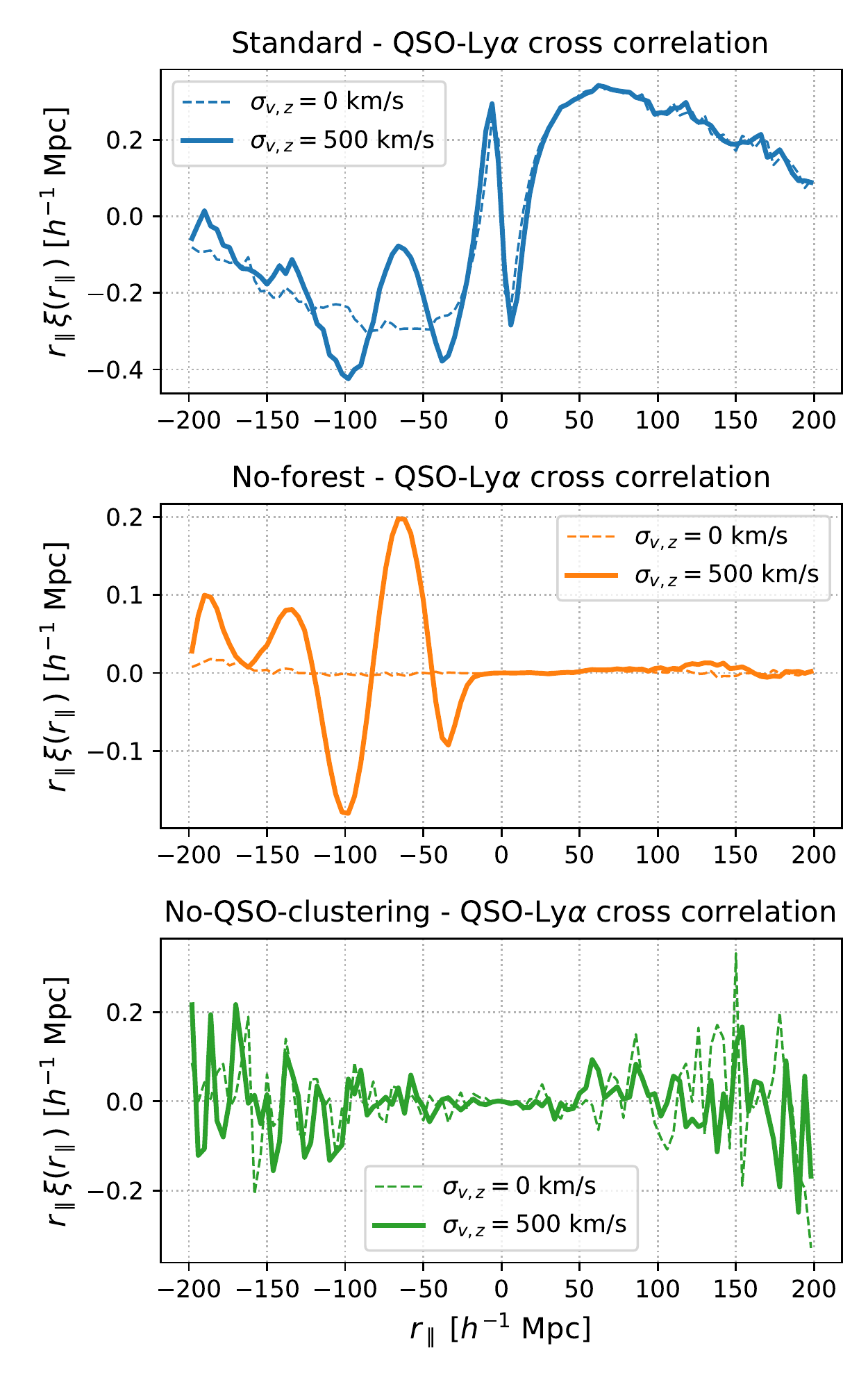}
    \includegraphics[width=0.45\textwidth]{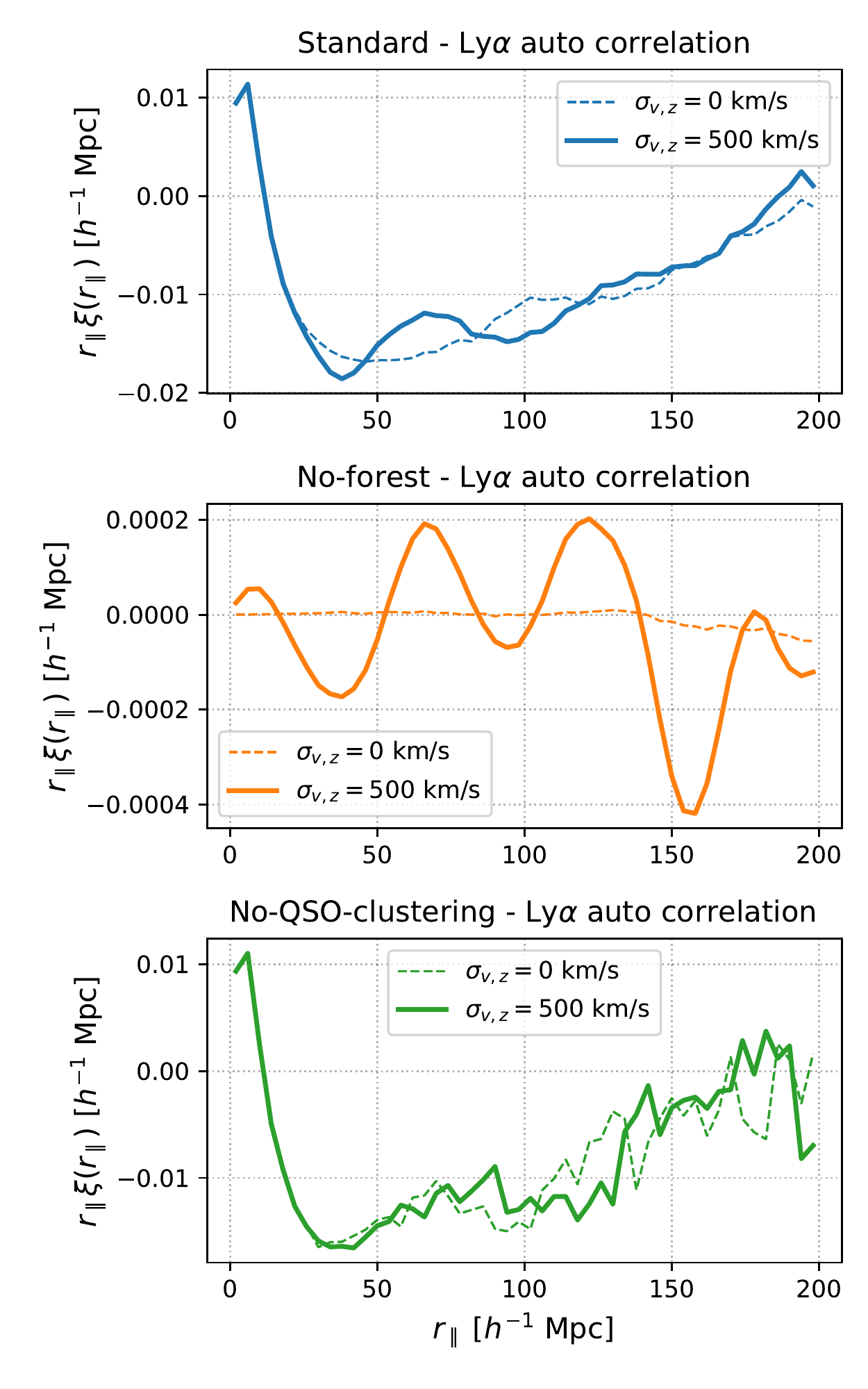}    
    \caption{Average correlation functions (left column: QSO x \lya cross-correlation, right column: \lya auto-correlation) at $r_\perp < 8$\hmpc as a function of radial separations $r_\parallel$ for standard mocks (blue), no-forest (orange) and no-QSO-clustering (green) mocks. Each panel shows the correlations for $\sigz = 0$ (dashed) and 500~\kms (solid). The oscillatory features caused by redshift errors can be seen in the standard and no-forest mocks, but as expected, they are not present in the no-QSO-clustering mocks.}
    
    \label{fig:slices}
\end{figure*}

The model for the contamination in the auto-correlation of \lya forests could be constructed as an extension of the cross-correlation model. However, results on mock catalogues show that the case of the auto-correlation is more complex than the cross correlation, indicating that there must be more ingredients to be taken into account in order to correctly reproduce the contamination. We leave this exploration for future work and we focus on results for the cross-correlation in the next section. 

\subsection{Contamination in special mocks}
\label{sec:model:ssec:special_mocks}

The contamination due to redshift errors on the correlations was further investigated by repeating the analysis on two special sets of mocks. These are the \emph{no-forest} and the \emph{no-QSO-clustering} mocks, described in detail in Section~\ref{sec:data:ssec:specialmocks}.
These are meant to test our hypothesis that the contamination arising from redshift errors depends on both the amplitude of the quasar clustering and the systematic error in the mean continuum estimate. This means that the effect should not depend in principle on the clustering of the \lya\ forest itself.

Figure~\ref{fig:slices} shows the average cross and auto-correlation functions at $r_\perp < 8$~\hmpc, plotted as functions of $r_\parallel$. Three sets of mocks are displayed: a stack of 10 realisations of standard mocks (blue), no-forest (orange) and a single realisation of no-QSO-clustering mocks (green). Each panel compares the radial correlations for mocks with and without redshift errors of $\sigz = 500$\kms. The oscillatory features caused by redshift errors are clearly seen in standard and no-forest mocks, while as expected, no-QSO-clustering mocks do not present any visible effect.

\begin{figure}
    \centering
    \includegraphics[width=0.45\textwidth]{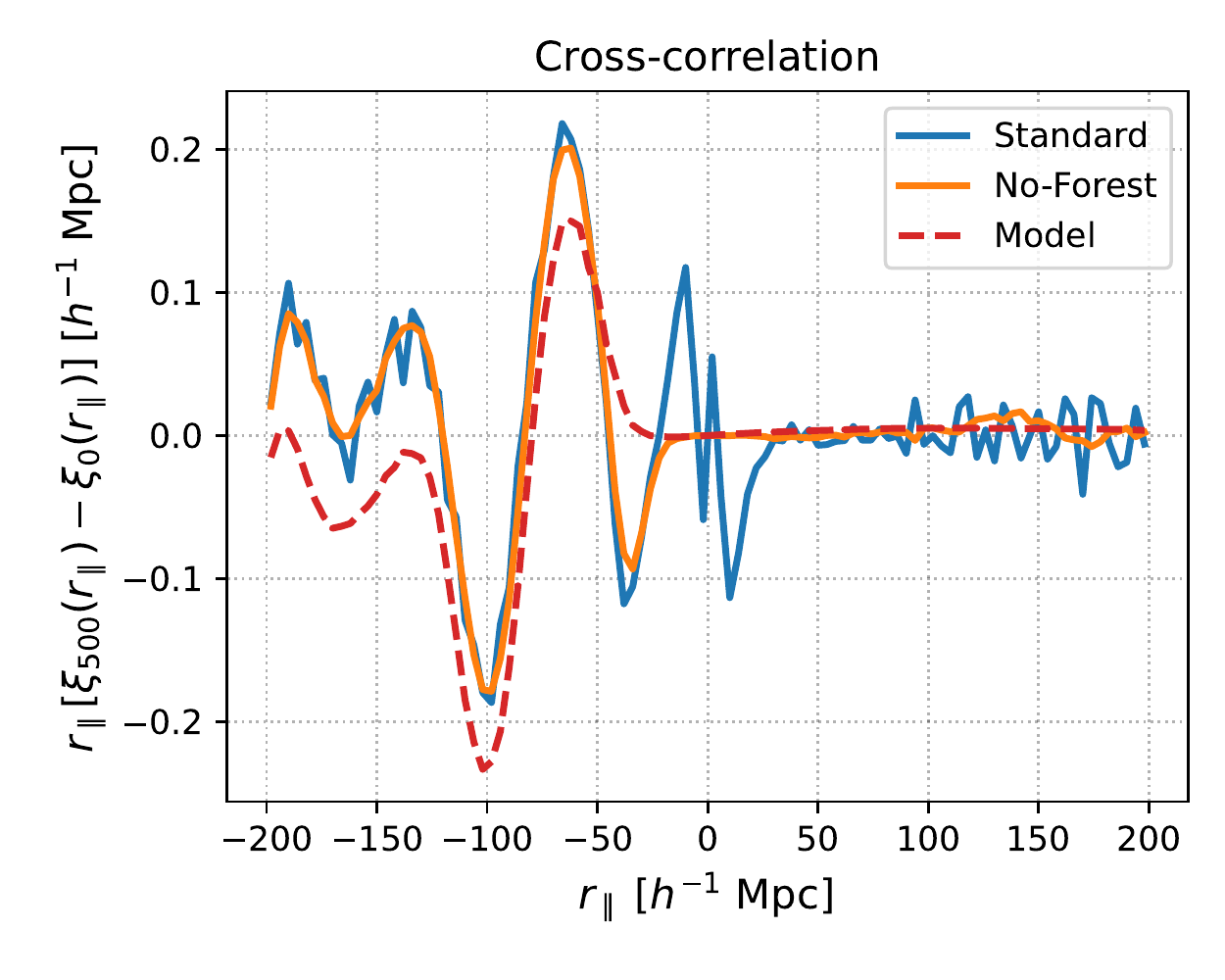}
    \caption{Cross-correlation functions at small transverse separations ($r_\perp < 8$\hmpc) as functions of radial separations $r_\parallel$ for mocks with $\sigz = 500$\kms minus the correlations when $\sigz = 0$. Two sets of mocks are shown: standard mocks in blue and no-forest in orange. The model derived in Section~\ref{sec:model}, indicated by the red dashed line, is in good qualitative agreement with the contamination by redshift errors.
    } 
    \label{fig:slices_model}
\end{figure}

Figure~\ref{fig:slices_model} displays the difference $\Delta \xi = \xi_{500} - \xi_{0}$ between the cross-correlations with and without errors. We observe a very good agreement between $\Delta \xi$ from standard and no-forest mocks, confirming our hypothesis.
Figure~\ref{fig:slices_model} also shows the $\Delta\xi = \xi_{500} - \xi_0$ for the model described in Section~\ref{sec:model} as a dashed red line. This model was computed using the bias in the mean continuum as shown in Fig.~\ref{fig:mean_cont} and the quasar auto-correlation displayed in Fig.~\ref{fig:auto_qq}.
In the case of the cross-correlation, the agreement between standard, no-forest and the model is very good, both in terms of the shape of the features and their amplitudes. 

The results on these special mocks and the qualitative agreement of our model for the contamination demonstrate that the effect of redshift errors in the cross-correlation depend both on the amplitude of the quasar clustering and on the systematic errors in the continuum shape.

\section{Discussion and Conclusions}
\label{sec:discussion}

In this paper, we have described the effect of errors in estimating quasar redshifts on the \lya\ forest correlation functions, and the consequent impact on the BAO parameters. We used mock \lya\ forests with redshifts of $z \in [2.1, 3.8]$ designed to simulate five years of the DESI program, to which we added various amplitudes of both Gaussian random peculiar velocities (Fingers of God) and Gaussian errors in the quasar redshift estimates. 

The two-point correlation functions exhibited unexpected systematic correlations at separations close to the line of sight when redshift errors were introduced (but which were absent when only large astrophysical FoG values were included). These features decrease in amplitude for increasing transverse separations, similar behaviour to the contamination caused by metals in the \lya\ forest. We found that these systematic correlations increase when increasing the amplitude of the redshift errors added to mocks. We believe it is the first time that this type of contamination has been observed and studied. 

We analysed the impact of such correlations on fits of the baryon acoustic oscillations. We found that BAO parameters are not significantly shifted by the addition of Gaussian random redshift errors, for the three cases: cross, auto, and joint fits of auto+cross-correlations. Redshift errors also cause the uncertainties in the BAO parameters to increase. We believe that if the model accounts for the systematic correlations it could reduce these uncertainties, which is important for current \lya\ surveys. These results are based on averages of ten realisations of the full 5-year DESI survey.

We derived a simplified model for the contamination to correlations from redshift errors, based on two main ingredients: the quasar auto-correlation function and the systematic bias in the mean continuum. These hypotheses were tested using special sets of mock catalogues, with either no-QSO-clustering or no \lya\ absorption. Mocks with no-QSO-clustering do not exhibit the features, while mocks with no-forest do contain them, confirming our hypothesis. The amplitudes and shapes of the contamination in these special mocks well describe the systematic correlations in the more realistic mock sets in the case of the cross-correlation between \lya\ and quasars. Modelling the contamination for the \lya\ auto-correlation is left for future work.

A detailed study of the effects analysed here for the first time is necessary to optimise the constraining power of the \lya\ forest sample of DESI. Mock catalogues will need to properly include redshift errors with more realistic models, not necessarily Gaussian, and consider how different emission lines have different velocity shifts \citep[e.g.,][]{hewett_2010, shen_sloan_2016}. The correlations between emission line velocity shifts could produce less smoothing of the forest continuum than implied by our prescription, which assumed random errors. This is because while the quasar redshift may be in error by some amount $\sigma_v$ relative to the systemic redshift, the quasar redshift determined from emission lines redward of Ly$\alpha$ may be a better predictor of the locations of the emission lines in the forest region than of the systemic redshift.

We did not study the issue of systematic biases in redshift estimates for quasars. Depending on which broad lines are present in the observed spectrum, systematic shifts may be introduced by spectral templates not accounting for them. The issue of systematic
biases, and how they impact BAO constraints has been studied in 
 \citet{font-riberaLargescaleQuasarLymanAlpha2013} and \citet{  glanvilleEffectSystematicRedshift2021}.

The methodology developed in this paper will be extremely useful in combination with a better understanding of the DESI data, to help develop fitting templates to account for these contaminations in the BAO analysis.

Finally, the contamination discussed in this paper could be an important systematic for studies that want to extract cosmological information from the full shape of the 3D correlations in the \lya\ forest \citep{Cuceu_2021_AP}. While the BAO measurements seem to be robust against them, it is possible that other cosmological inference might be biased if the impact of redshift errors is not taken into account.

\section*{Acknowledgements}

SY was supported by a Science and Technology Facilities Council (STFC) studentship. The project leading to this publication has received funding from Excellence Initiative of Aix-Marseille University - A*MIDEX, a French ''Investissements d'Avenir'' programme (AMX-20-CE-02 - DARKUNI). AFR acknowledges funds through the program Ramon y Cajal (RYC-2018-025210) of the Spanish Ministry of Science and Innovation. IFAE is partially funded by the CERCA program of the Generalitat de Catalunya. IPR was supported by funding from the European Union's Horizon 2020 research and innovation programme under the Marie Skłodowska-Curie grant agreement No. 754510.

 This research is supported by the Director, Office of Science, Office of High Energy Physics of the U.S. Department of Energy under Contract No. DE–AC02–05CH11231, and by the National Energy Research Scientific Computing Center, a DOE Office of Science User Facility under the same contract; additional support for DESI is provided by the U.S. National Science Foundation, Division of Astronomical Sciences under Contract No. AST-0950945 to the NSF’s National Optical-Infrared Astronomy Research Laboratory; the Science and Technologies Facilities Council of the United Kingdom; the Gordon and Betty Moore Foundation; the Heising-Simons Foundation; the French Alternative Energies and Atomic Energy Commission (CEA); the National Council of Science and Technology of Mexico (CONACYT); the Ministry of Science and Innovation of Spain (MICINN), and by the DESI Member Institutions: https://www.desi.lbl.gov/collaborating-institutions.

The authors are honored to be permitted to conduct scientific research on Iolkam Du’ag (Kitt Peak), a mountain with particular significance to the Tohono O’odham Nation. 

This research used resources of the National Energy Research Scientific Computing Center (NERSC), a U.S. Department of Energy Office of Science User Facility located at Lawrence Berkeley National Laboratory, operated under Contract No. DE-AC02-05CH11231.

\section*{Data Availability}

The analysis code \textsc{picca} is publicly available at \url{https://github.com/igmhub/picca}.

Data points for all plots are available at \url{https://zenodo.org/record/6543559}.


\bibliographystyle{mnras}
\bibliography{mnras_template}

\begin{thebibliography}{}
\makeatletter
\relax
\def\mn@urlcharsother{\let\do\@makeother \do\$\do\&\do\#\do\^\do\_\do\%\do\~}
\def\mn@doi{\begingroup\mn@urlcharsother \@ifnextchar [ {\mn@doi@}
  {\mn@doi@[]}}
\def\mn@doi@[#1]#2{\def\@tempa{#1}\ifx\@tempa\@empty \href
  {http://dx.doi.org/#2} {doi:#2}\else \href {http://dx.doi.org/#2} {#1}\fi
  \endgroup}
\def\mn@eprint#1#2{\mn@eprint@#1:#2::\@nil}
\def\mn@eprint@arXiv#1{\href {http://arxiv.org/abs/#1} {{\tt arXiv:#1}}}
\def\mn@eprint@dblp#1{\href {http://dblp.uni-trier.de/rec/bibtex/#1.xml}
  {dblp:#1}}
\def\mn@eprint@#1:#2:#3:#4\@nil{\def\@tempa {#1}\def\@tempb {#2}\def\@tempc
  {#3}\ifx \@tempc \@empty \let \@tempc \@tempb \let \@tempb \@tempa \fi \ifx
  \@tempb \@empty \def\@tempb {arXiv}\fi \@ifundefined
  {mn@eprint@\@tempb}{\@tempb:\@tempc}{\expandafter \expandafter \csname
  mn@eprint@\@tempb\endcsname \expandafter{\@tempc}}}

\bibitem[\protect\citeauthoryear{Ahumada et~al.,}{Ahumada
  et~al.}{2020}]{ahumada16thDataRelease2020}
Ahumada R.,  et~al., 2020, \mn@doi [The Astrophysical Journal Supplement
  Series] {10.3847/1538-4365/ab929e}, 249, 3

\bibitem[\protect\citeauthoryear{Alam et~al.,}{Alam
  et~al.}{2021}]{alam_completed_2021}
Alam S.,  et~al., 2021, \mn@doi [Physical Review D]
  {10.1103/PhysRevD.103.083533}, 103, 083533

\bibitem[\protect\citeauthoryear{Bautista et~al.,}{Bautista
  et~al.}{2015}]{bautista_mocks_2015}
Bautista J.~E.,  et~al., 2015, \mn@doi [Journal of Cosmology and Astroparticle
  Physics] {10.1088/1475-7516/2015/05/060}, 2015, 060

\bibitem[\protect\citeauthoryear{Bautista et~al.,}{Bautista
  et~al.}{2017}]{bautista_measurement_2017}
Bautista J.~E.,  et~al., 2017, \mn@doi [Astronomy \& Astrophysics]
  {10.1051/0004-6361/201730533}, 603, A12

\bibitem[\protect\citeauthoryear{{Blomqvist} et~al.,}{{Blomqvist}
  et~al.}{2019}]{Blomqvist:2019}
{Blomqvist} M.,  et~al., 2019, \mn@doi [\aap] {10.1051/0004-6361/201935641},
  \href {https://ui.adsabs.harvard.edu/abs/2019A&A...629A..86B} {629, A86}

\bibitem[\protect\citeauthoryear{Borde, Palanque-Delabrouille, Rossi, Viel,
  Bolton, Y{\`{e}}che, LeGoff  \& Rich}{Borde
  et~al.}{2014}]{borde_hydrosim_2014}
Borde A.,  Palanque-Delabrouille N.,  Rossi G.,  Viel M.,  Bolton J.~S.,
  Y{\`{e}}che C.,  LeGoff J.-M.,   Rich J.,  2014, \mn@doi [Journal of
  Cosmology and Astroparticle Physics] {10.1088/1475-7516/2014/07/005}, 2014,
  005

\bibitem[\protect\citeauthoryear{Bose, Eisenstein, Hadzhiyska, Garrison  \&
  Yuan}{Bose et~al.}{2021}]{Bose_AbacusSummit:2021}
Bose S.,  Eisenstein D.~J.,  Hadzhiyska B.,  Garrison L.~H.,   Yuan S.,  2021,
  Constructing high-fidelity halo merger trees in AbacusSummit (\mn@eprint
  {arXiv} {2110.11409})

\bibitem[\protect\citeauthoryear{Brout et~al.,}{Brout
  et~al.}{2019}]{brout_des_2019}
Brout D.,  et~al., 2019, \mn@doi [The Astrophysical Journal]
  {10.3847/1538-4357/ab08a0}, 874, 150

\bibitem[\protect\citeauthoryear{Busca \& Balland}{Busca \&
  Balland}{2018}]{busca_quasarnet:_2018}
Busca N.,  Balland C.,  2018, arXiv:1808.09955 [astro-ph]

\bibitem[\protect\citeauthoryear{Busca et~al.,}{Busca
  et~al.}{2013}]{busca_baryon_2013}
Busca N.~G.,  et~al., 2013, \mn@doi [Astronomy \& Astrophysics]
  {10.1051/0004-6361/201220724}, 552, A96

\bibitem[\protect\citeauthoryear{Chabanier, Bournaud, Dubois,
  Palanque-Delabrouille, Y{\`{e}}che, Armengaud, Peirani  \&
  Beckmann}{Chabanier et~al.}{2020}]{chabanier_AGN_2020}
Chabanier S.,  Bournaud F.,  Dubois Y.,  Palanque-Delabrouille N.,  Y{\`{e}}che
  C.,  Armengaud E.,  Peirani S.,   Beckmann R.~S.,  2020, arXiv:2002.02822v3
  [astro-ph]

\bibitem[\protect\citeauthoryear{{Chaves-Montero}, {Angulo}  \&
  {Hern{\'a}ndez-Monteagudo}}{{Chaves-Montero}
  et~al.}{2018}]{Chaves-Montero_2018}
{Chaves-Montero} J.,  {Angulo} R.~E.,   {Hern{\'a}ndez-Monteagudo} C.,  2018,
  \mn@doi [\mnras] {10.1093/mnras/sty924}, \href
  {https://ui.adsabs.harvard.edu/abs/2018MNRAS.477.3892C} {477, 3892}

\bibitem[\protect\citeauthoryear{Cole et~al.,}{Cole
  et~al.}{2005}]{cole_2df_2005}
Cole S.,  et~al., 2005, \mn@doi [Monthly Notices of the Royal Astronomical
  Society] {10.1111/j.1365-2966.2005.09318.x}, 362, 505

\bibitem[\protect\citeauthoryear{Coles \& Jones}{Coles \&
  Jones}{1991}]{coles_lognormal_1991}
Coles P.,  Jones B.,  1991, \mn@doi [\mnras] {10.1093/mnras/248.1.1}, \href
  {https://ui.adsabs.harvard.edu/abs/1991MNRAS.248....1C} {248, 1}

\bibitem[\protect\citeauthoryear{{Cuceu}, {Font-Ribera}, {Joachimi}  \&
  {Nadathur}}{{Cuceu} et~al.}{2021}]{Cuceu_2021_AP}
{Cuceu} A.,  {Font-Ribera} A.,  {Joachimi} B.,   {Nadathur} S.,  2021, \mn@doi
  [\mnras] {10.1093/mnras/stab1999}, \href
  {https://ui.adsabs.harvard.edu/abs/2021MNRAS.506.5439C} {506, 5439}

\bibitem[\protect\citeauthoryear{{DESI Collaboration} et~al.,}{{DESI
  Collaboration} et~al.}{2016}]{desi_collaboration_desi_2016}
{DESI Collaboration} et~al., 2016, arXiv:1611.00036 [astro-ph]

\bibitem[\protect\citeauthoryear{\DE{Sainte Agathe}{de}{de} Sainte~Agathe
  et~al.,}{\DE{Sainte Agathe}{de}{de} Sainte~Agathe
  et~al.}{2019}]{desainteagatheBaryonAcousticOscillations2019}
\DE{Sainte Agathe}{de}{de} Sainte~Agathe V.,  et~al., 2019, \mn@doi [\aap]
  {10.1051/0004-6361/201935638}, \href
  {https://ui.adsabs.harvard.edu/abs/2019A&A...629A..85D} {629, A85}

\bibitem[\protect\citeauthoryear{\DU{Mas des Bourboux}{du}{du} Mas~des Bourboux
  et~al.,}{\DU{Mas des Bourboux}{du}{du} Mas~des Bourboux
  et~al.}{2017}]{duMasDesBourboux:2017}
\DU{Mas des Bourboux}{du}{du} Mas~des Bourboux H.,  et~al., 2017, Astronomy and
  Astrophysics, 608, A130

\bibitem[\protect\citeauthoryear{\DU{Mas des Bourboux}{du}{du} Mas~des Bourboux
  et~al.,}{\DU{Mas des Bourboux}{du}{du} Mas~des Bourboux
  et~al.}{2019}]{du_Mas_des_Bourboux:2019}
\DU{Mas des Bourboux}{du}{du} Mas~des Bourboux H.,  et~al., 2019, \mn@doi [The
  Astrophysical Journal] {10.3847/1538-4357/ab1d49}, 878, 47

\bibitem[\protect\citeauthoryear{\DU{Mas des Bourboux}{du}{du} Mas~des Bourboux
  et~al.,}{\DU{Mas des Bourboux}{du}{du} Mas~des Bourboux
  et~al.}{2020}]{du_mas_des_bourboux_completed_2020}
\DU{Mas des Bourboux}{du}{du} Mas~des Bourboux H.,  et~al., 2020, \mn@doi [The
  Astrophysical Journal] {10.3847/1538-4357/abb085}, 901, 153

\bibitem[\protect\citeauthoryear{\DU{Mas des Bourboux}{du}{du} Mas~des Bourboux
  et~al.,}{\DU{Mas des Bourboux}{du}{du} Mas~des Bourboux
  et~al.}{2021}]{du_mas_des_bourboux_picca_2021}
\DU{Mas des Bourboux}{du}{du} Mas~des Bourboux H.,  et~al., 2021, {picca:
  Package for Igm Cosmological-Correlations Analyses} (\mn@eprint {ascl}
  {2106.018})

\bibitem[\protect\citeauthoryear{Dawson et~al.,}{Dawson
  et~al.}{2016}]{dawson_extended_2016}
Dawson K.~S.,  et~al., 2016, \mn@doi [The Astronomical Journal]
  {10.3847/0004-6256/151/2/44}, 151, 44

\bibitem[\protect\citeauthoryear{Delubac et~al.,}{Delubac
  et~al.}{2015}]{delubac_baryon_2015}
Delubac T.,  et~al., 2015, \mn@doi [Astronomy \& Astrophysics]
  {10.1051/0004-6361/201423969}, 574, A59

\bibitem[\protect\citeauthoryear{Eisenstein et~al.,}{Eisenstein
  et~al.}{2005}]{eisenstein_detection_2005}
Eisenstein D.~J.,  et~al., 2005, \mn@doi [The Astrophysical Journal]
  {10.1086/466512}, 633, 560

\bibitem[\protect\citeauthoryear{Farr et~al.,}{Farr
  et~al.}{2020a}]{farr_lyacolore_2020}
Farr J.,  et~al., 2020a, \mn@doi [Journal of Cosmology and Astroparticle
  Physics] {10.1088/1475-7516/2020/03/068}, 2020, 068

\bibitem[\protect\citeauthoryear{{Farr}, {Font-Ribera}  \& {Pontzen}}{{Farr}
  et~al.}{2020b}]{farr:2020_quasarnet}
{Farr} J.,  {Font-Ribera} A.,   {Pontzen} A.,  2020b, \mn@doi [\jcap]
  {10.1088/1475-7516/2020/11/015}, \href
  {https://ui.adsabs.harvard.edu/abs/2020JCAP...11..015F} {2020, 015}

\bibitem[\protect\citeauthoryear{{Font-Ribera} \&
  {Miralda-Escud{\'e}}}{{Font-Ribera} \&
  {Miralda-Escud{\'e}}}{2012}]{FontRibera_2012_HCD}
{Font-Ribera} A.,  {Miralda-Escud{\'e}} J.,  2012, \mn@doi [\jcap]
  {10.1088/1475-7516/2012/07/028}, \href
  {https://ui.adsabs.harvard.edu/abs/2012JCAP...07..028F} {2012, 028}

\bibitem[\protect\citeauthoryear{{Font-Ribera}, {McDonald}  \&
  {Miralda-Escud{\'e}}}{{Font-Ribera} et~al.}{2012a}]{FontRibera_2012_mocks}
{Font-Ribera} A.,  {McDonald} P.,   {Miralda-Escud{\'e}} J.,  2012a, \mn@doi
  [\jcap] {10.1088/1475-7516/2012/01/001}, \href
  {https://ui.adsabs.harvard.edu/abs/2012JCAP...01..001F} {2012, 001}

\bibitem[\protect\citeauthoryear{{Font-Ribera} et~al.,}{{Font-Ribera}
  et~al.}{2012b}]{FontRibera_2012_DLA_cross}
{Font-Ribera} A.,  et~al., 2012b, \mn@doi [\jcap]
  {10.1088/1475-7516/2012/11/059}, \href
  {https://ui.adsabs.harvard.edu/abs/2012JCAP...11..059F} {2012, 059}

\bibitem[\protect\citeauthoryear{{Font-Ribera} et~al.,}{{Font-Ribera}
  et~al.}{2013}]{font-riberaLargescaleQuasarLymanAlpha2013}
{Font-Ribera} A.,  et~al., 2013, \mn@doi [\jcap]
  {10.1088/1475-7516/2013/05/018}, \href
  {https://ui.adsabs.harvard.edu/abs/2013JCAP...05..018F} {2013, 018}

\bibitem[\protect\citeauthoryear{{Font-Ribera} et~al.,}{{Font-Ribera}
  et~al.}{2014}]{Font-Ribera:2014_xcf}
{Font-Ribera} A.,  et~al., 2014, \mn@doi [\jcap]
  {10.1088/1475-7516/2014/05/027}, \href
  {https://ui.adsabs.harvard.edu/abs/2014JCAP...05..027F} {2014, 027}

\bibitem[\protect\citeauthoryear{Garrison, Eisenstein, Ferrer, Maksimova  \&
  Pinto}{Garrison et~al.}{2021}]{Garrison_AbacusSummit:2021}
Garrison L.~H.,  Eisenstein D.~J.,  Ferrer D.,  Maksimova N.~A.,   Pinto P.~A.,
   2021, \mn@doi [Monthly Notices of the Royal Astronomical Society]
  {10.1093/mnras/stab2482}, 508, 575

\bibitem[\protect\citeauthoryear{{Glanville}, {Howlett}  \&
  {Davis}}{{Glanville} et~al.}{2021}]{glanvilleEffectSystematicRedshift2021}
{Glanville} A.,  {Howlett} C.,   {Davis} T.~M.,  2021, \mn@doi [\mnras]
  {10.1093/mnras/stab657}, \href
  {https://ui.adsabs.harvard.edu/abs/2021MNRAS.503.3510G} {503, 3510}

\bibitem[\protect\citeauthoryear{Hadzhiyska, Garrison, Eisenstein  \&
  Bose}{Hadzhiyska et~al.}{2021}]{Hadzhiyska_AbacusSummit:2021}
Hadzhiyska B.,  Garrison L.~H.,  Eisenstein D.,   Bose S.,  2021, \mn@doi
  [Monthly Notices of the Royal Astronomical Society] {10.1093/mnras/stab3066}

\bibitem[\protect\citeauthoryear{{Harris} et~al.,}{{Harris}
  et~al.}{2016}]{harrisCompositeSpectrumBOSS2016}
{Harris} D.~W.,  et~al., 2016, \mn@doi [\aj] {10.3847/0004-6256/151/6/155},
  \href {https://ui.adsabs.harvard.edu/abs/2016AJ....151..155H} {151, 155}

\bibitem[\protect\citeauthoryear{Hewett \& Wild}{Hewett \&
  Wild}{2010}]{hewett_2010}
Hewett P.~C.,  Wild V.,  2010, \mn@doi [Monthly Notices of the Royal
  Astronomical Society] {10.1111/j.1365-2966.2010.16648.x}, 405, 2302

\bibitem[\protect\citeauthoryear{Kirkby et~al.,}{Kirkby
  et~al.}{2013}]{kirkby_fitting_2013}
Kirkby D.,  et~al., 2013, \mn@doi [Journal of Cosmology and Astroparticle
  Physics] {10.1088/1475-7516/2013/03/024}, 2013, 024

\bibitem[\protect\citeauthoryear{{Landy} \& {Szalay}}{{Landy} \&
  {Szalay}}{1993}]{landy_estimator_1993}
{Landy} S.~D.,  {Szalay} A.~S.,  1993, \mn@doi [\apj] {10.1086/172900}, \href
  {https://ui.adsabs.harvard.edu/abs/1993ApJ...412...64L} {412, 64}

\bibitem[\protect\citeauthoryear{Le~Goff et~al.,}{Le~Goff
  et~al.}{2011}]{le_goff_simulations_2011}
Le~Goff J.~M.,  et~al., 2011, \mn@doi [Astronomy \& Astrophysics]
  {10.1051/0004-6361/201117736}, 534, A135

\bibitem[\protect\citeauthoryear{{Lewis}, {Challinor}  \& {Lasenby}}{{Lewis}
  et~al.}{2000}]{lewis_camb_2000}
{Lewis} A.,  {Challinor} A.,   {Lasenby} A.,  2000, \mn@doi [\apj]
  {10.1086/309179}, \href
  {https://ui.adsabs.harvard.edu/abs/2000ApJ...538..473L} {538, 473}

\bibitem[\protect\citeauthoryear{Lyke et~al.,}{Lyke
  et~al.}{2020}]{lyke_sloan_2020}
Lyke B.~W.,  et~al., 2020, \mn@doi [The Astrophysical Journal Supplement
  Series] {10.3847/1538-4365/aba623}, 250, 8

\bibitem[\protect\citeauthoryear{Maksimova, Garrison, Eisenstein, Hadzhiyska,
  Bose  \& Satterthwaite}{Maksimova et~al.}{2021}]{Maksimova_AbacusSummit:2021}
Maksimova N.~A.,  Garrison L.~H.,  Eisenstein D.~J.,  Hadzhiyska B.,  Bose S.,
   Satterthwaite T.~P.,  2021, Monthly Notices of the Royal Astronomical
  Society, 508, 4017

\bibitem[\protect\citeauthoryear{{McGreer}, {Moustakas}  \&
  {Schindler}}{{McGreer} et~al.}{2021}]{2021ascl.soft06008M}
{McGreer} I.,  {Moustakas} J.,   {Schindler} J.,  2021, {simqso: Simulated
  quasar spectra generator} (\mn@eprint {ascl} {2106.008})

\bibitem[\protect\citeauthoryear{Peirani, Weinberg, Colombi, Blaizot, Dubois
  \& Pichon}{Peirani et~al.}{2014}]{peirani_lymas_2014}
Peirani S.,  Weinberg D.~H.,  Colombi S.,  Blaizot J.,  Dubois Y.,   Pichon C.,
   2014, \mn@doi [The Astrophysical Journal] {10.1088/0004-637x/784/1/11}, 784,
  11

\bibitem[\protect\citeauthoryear{{Percival} \& {White}}{{Percival} \&
  {White}}{2009}]{percival_white_2009}
{Percival} W.~J.,  {White} M.,  2009, \mn@doi [\mnras]
  {10.1111/j.1365-2966.2008.14211.x}, \href
  {https://ui.adsabs.harvard.edu/abs/2009MNRAS.393..297P} {393, 297}

\bibitem[\protect\citeauthoryear{Perlmutter et~al.,}{Perlmutter
  et~al.}{1998}]{perlmutter_discovery_1998}
Perlmutter S.,  et~al., 1998, \mn@doi [Nature] {10.1038/34124}, 391, 51

\bibitem[\protect\citeauthoryear{Pérez-Ràfols, Pieri, Blomqvist, Morrison  \&
  Som}{Pérez-Ràfols et~al.}{2020}]{perez-rafols_spectroscopic_2020}
Pérez-Ràfols I.,  Pieri M.~M.,  Blomqvist M.,  Morrison S.,   Som D.,  2020,
  \mn@doi [Monthly Notices of the Royal Astronomical Society]
  {10.1093/mnras/stz3467}, 496, 4931

\bibitem[\protect\citeauthoryear{Ramírez-Pérez, Sanchez, Alonso  \&
  Font-Ribera}{Ramírez-Pérez et~al.}{2021}]{Ramirez-Perez:2021}
Ramírez-Pérez C.,  Sanchez J.,  Alonso D.,   Font-Ribera A.,  2021,
  arXiv:2111.05069v1 [astro-ph]

\bibitem[\protect\citeauthoryear{Riess et~al.,}{Riess
  et~al.}{1998}]{riess_observational_1998}
Riess A.~G.,  et~al., 1998, \mn@doi [The Astronomical Journal]
  {10.1086/300499}, 116, 1009

\bibitem[\protect\citeauthoryear{Rossi, Palanque-Delabrouille, Borde, Viel,
  Y\`eche, Bolton, Rich  \& Goff}{Rossi et~al.}{2014}]{rossi_hydrosim_2014}
Rossi G.,  Palanque-Delabrouille N.,  Borde A.,  Viel M.,  Y\`eche C.,  Bolton
  J.~S.,  Rich J.,   Goff J.-M.~L.,  2014, \mn@doi [A\&A]
  {10.1051/0004-6361/201423507}, 567, A79

\bibitem[\protect\citeauthoryear{Scolnic et~al.,}{Scolnic
  et~al.}{2018}]{scolnic_complete_2018}
Scolnic D.~M.,  et~al., 2018, \mn@doi [The Astrophysical Journal]
  {10.3847/1538-4357/aab9bb}, 859, 101

\bibitem[\protect\citeauthoryear{Shen et~al.,}{Shen
  et~al.}{2016}]{shen_sloan_2016}
Shen Y.,  et~al., 2016, \mn@doi [The Astrophysical Journal]
  {10.3847/0004-637X/831/1/7}, 831, 7

\bibitem[\protect\citeauthoryear{Slosar et~al.,}{Slosar
  et~al.}{2013}]{slosar_measurement_2013}
Slosar A.,  et~al., 2013, \mn@doi [Journal of Cosmology and Astroparticle
  Physics] {10.1088/1475-7516/2013/04/026}, 2013, 026

\bibitem[\protect\citeauthoryear{Sorini, O{\~{n}}orbe, Luki{\'{c}}  \&
  Hennawi}{Sorini et~al.}{2016}]{sorini_sims_2016}
Sorini D.,  O{\~{n}}orbe J.,  Luki{\'{c}} Z.,   Hennawi J.~F.,  2016, \mn@doi
  [The Astrophysical Journal] {10.3847/0004-637x/827/2/97}, 827, 97

\bibitem[\protect\citeauthoryear{Walther, Armengaud, Ravoux,
  Palanque-Delabrouille, Y{\`{e}}che  \& Luki{\'{c}}}{Walther
  et~al.}{2021}]{walther_nyx_2021}
Walther M.,  Armengaud E.,  Ravoux C.,  Palanque-Delabrouille N.,  Y{\`{e}}che
  C.,   Luki{\'{c}} Z.,  2021, \mn@doi [Journal of Cosmology and Astroparticle
  Physics] {10.1088/1475-7516/2021/04/059}, 2021, 059

\makeatother
\end{thebibliography}



\appendix


\bsp	
\label{lastpage}
\end{document}